\documentclass[sigconf, screen]{acmart}

\copyrightyear{2018}
\acmYear{2018}
\setcopyright{acmcopyright}
\acmConference[MM '18]{2018 ACM Multimedia Conference}{October 22--26,
2018}{Seoul, Republic of Korea}
\acmBooktitle{2018 ACM Multimedia Conference (MM '18), October 22--26,
2018, Seoul, Republic of Korea}
\acmPrice{15.00}
\acmDOI{10.1145/3240508.3240545}
\acmISBN{978-1-4503-5665-7/18/10}
\usepackage{flushend,cuted}
\usepackage{graphicx}
\usepackage{subfigure}
\usepackage{booktabs} 
\usepackage{multirow}
\usepackage[T1]{fontenc}
\setcopyright{acmcopyright}
\acmArticle{4}
\acmPrice{15.00}

\begin{document}
\title[QARC: Video Quality Aware Rate Control]{QARC: Video Quality Aware Rate Control for Real-Time Video Streaming via Deep Reinforcement Learning}


\author{Tianchi Huang$^{3,1}$, Rui-Xiao Zhang$^{1}$, Chao Zhou$^{2*}$, Lifeng Sun$^{1*}$}
\affiliation{$^{1}$ Dept. of Computer Science and Technology, Tsinghua University, Beijing, China}
\affiliation{$^{2}$ Beijing Kuaishou Technology Co., Ltd., China}
\affiliation{$^{3}$ Dept. of Computer Science and Technology, Guizhou University, Guizhou, China}
\affiliation{\{htc17,zhangrx17\}@mails.tsinghua.edu.cn, zhouchao@kuaishou.com, sunlf@tsinghua.edu.cn}
\renewcommand{\shortauthors}{Huang et al.}
\renewcommand{\authors}{Tianchi Huang, Rui-Xiao Zhang, Chao Zhou, Lifeng Sun}
\begin{abstract}
Real-time video streaming is now one of the main applications in all network environments. Due to the fluctuation of throughput under various network conditions, how to choose a proper bitrate adaptively has become an upcoming and interesting issue. To tackle this problem, most proposed rate control methods work for providing high video bitrates instead of video qualities. Nevertheless, we notice that there exists a trade-off between sending bitrate and video quality, which motivates us to focus on how to reach a balance between them. 

In this paper, we propose QARC (video Quality Aware Rate Control), a rate control algorithm that aims to obtain a higher perceptual video quality with possible lower sending rate and transmission latency. Starting from scratch, QARC uses deep reinforcement learning(DRL) algorithm to train a neural network for selecting future bitrates based on previously observed network status and past video frames. To overcome the ``state explosion problem'', we design a neural network to predict future perceptual video quality as a vector for taking the place of the raw picture in the DRL's inputs. 

We evaluate QARC via trace-driven simulation, outperforming existing approach with improvements in average video quality of 18\% - 25\% and decreasing in average latency with 23\% -45\%. Meanwhile, comparing QARC with offline optimal high bitrate method on various network conditions, we find that QARC also yields a solid result.

\end{abstract}

\maketitle

\section{Introduction}
\begin{figure}
  \centering
  \begin{minipage}{1.0\linewidth}
      \centering 
      \subfigure[A sample video clip with static video background~\cite{beyourself}]{\includegraphics[width=0.5\textwidth]{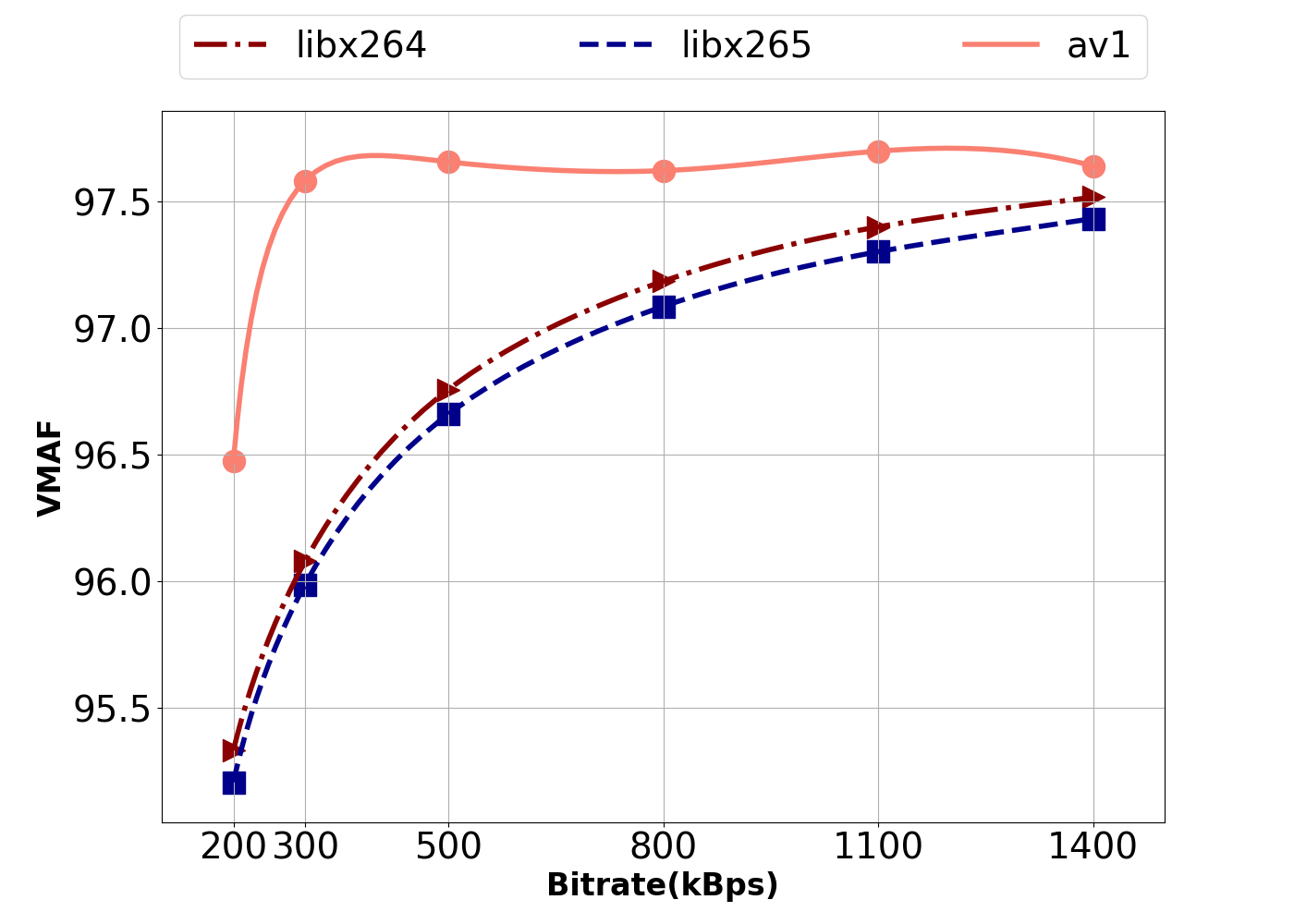}
 \includegraphics[width=0.5\textwidth]{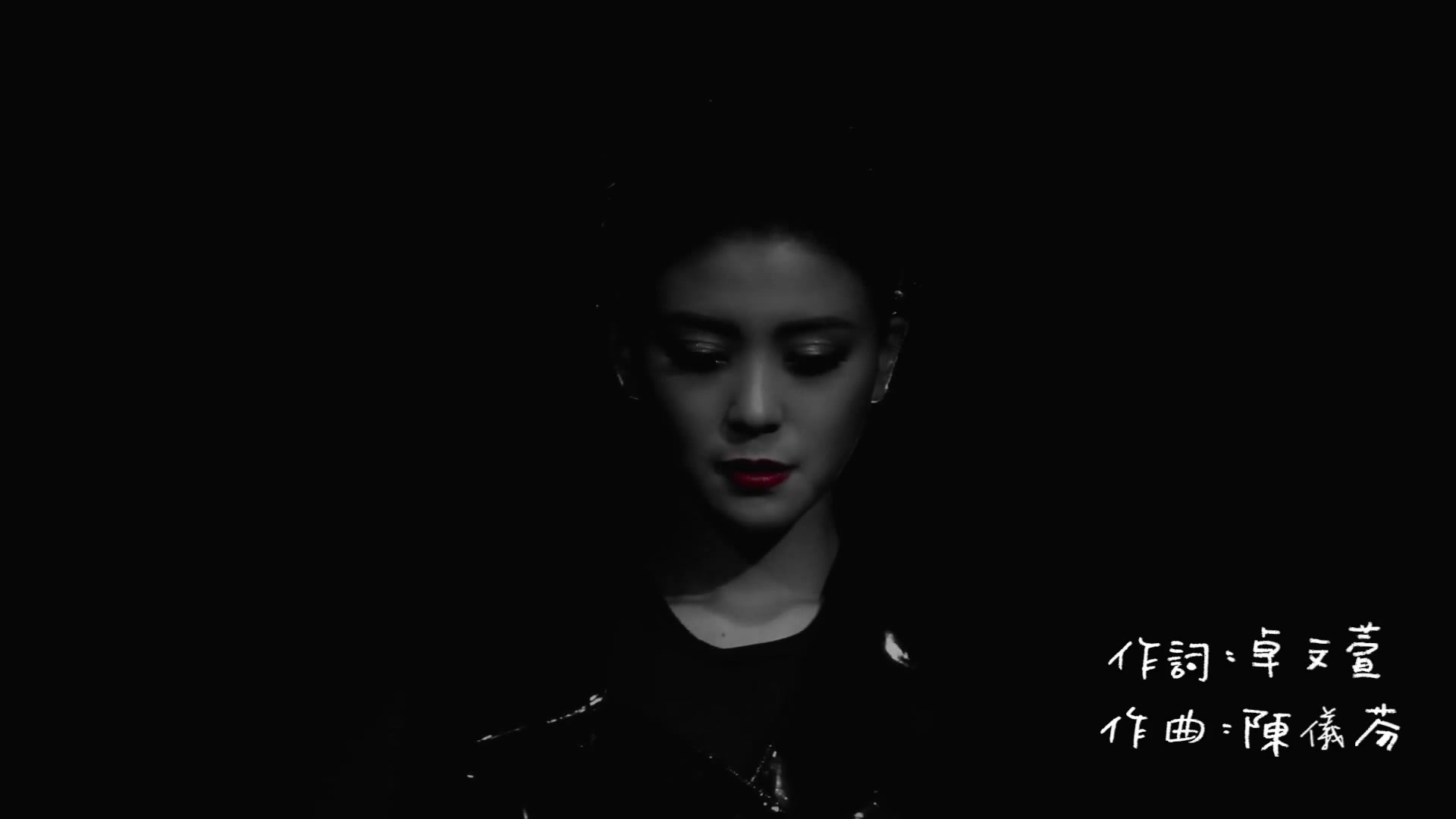}}
  \end{minipage}
  \begin{minipage}{1.0\linewidth}
      \centering
      \subfigure[A sample video clip with dynamic video scene\cite{KiboutekiRefrain}]{\includegraphics[width=0.5\textwidth]{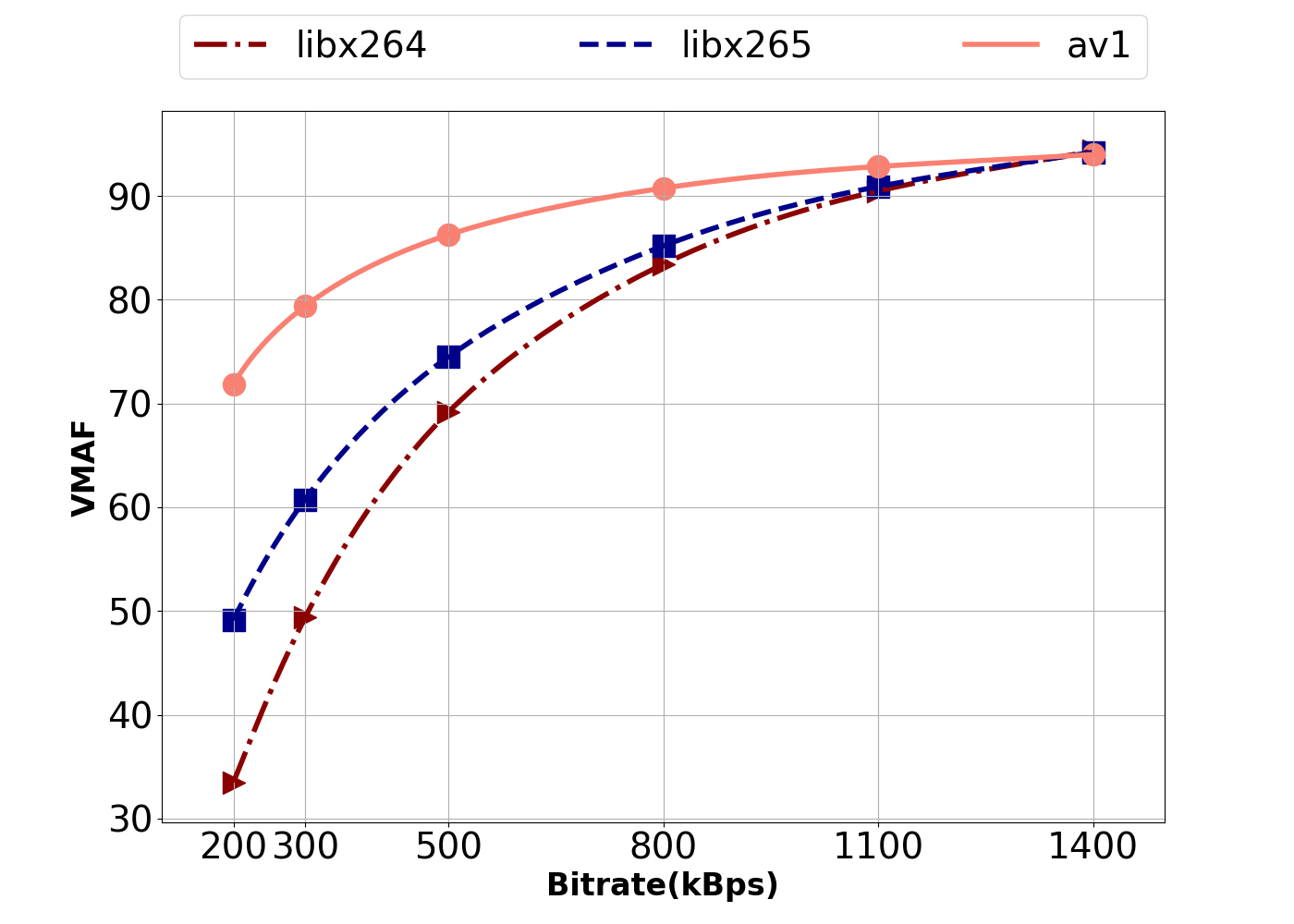}
 \includegraphics[width=0.5\textwidth]{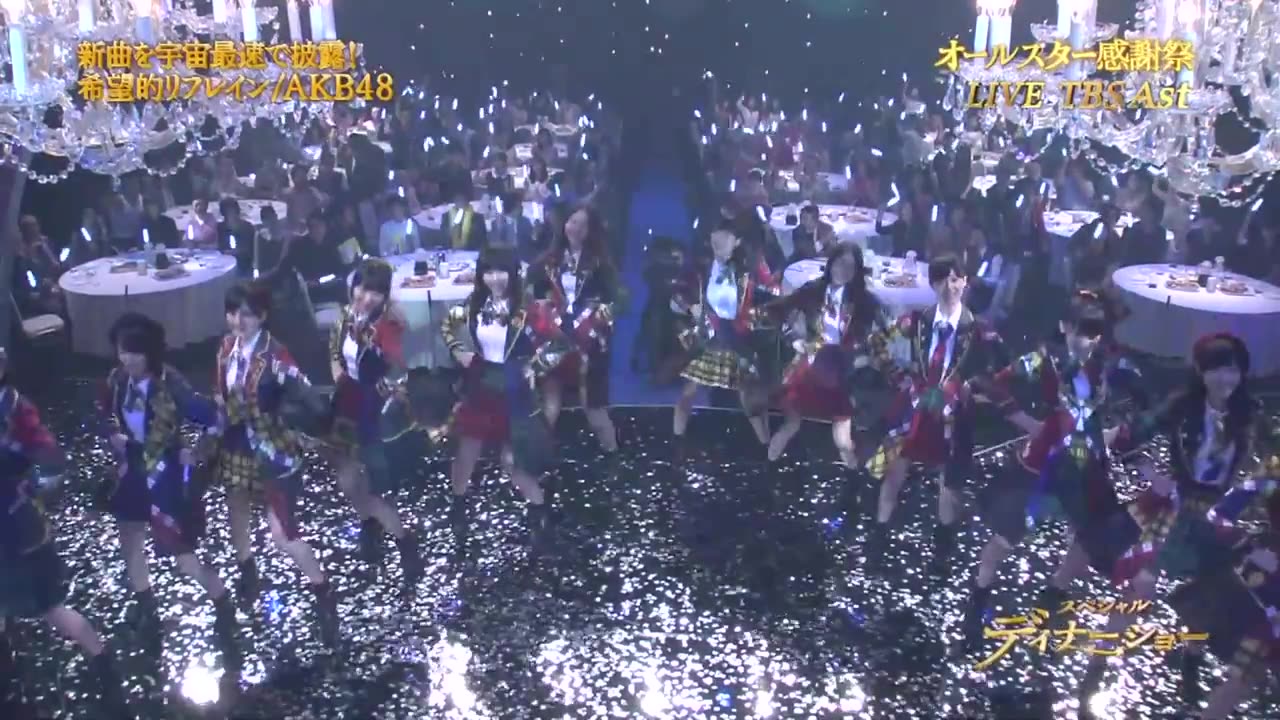}}
  \end{minipage}  
  \begin{minipage}{1.0\linewidth}
      \centering 
      \subfigure[A sample video clip with both static and dynamic video scene\cite{ILoveIEmbrace}]{\includegraphics[width=0.5\textwidth]{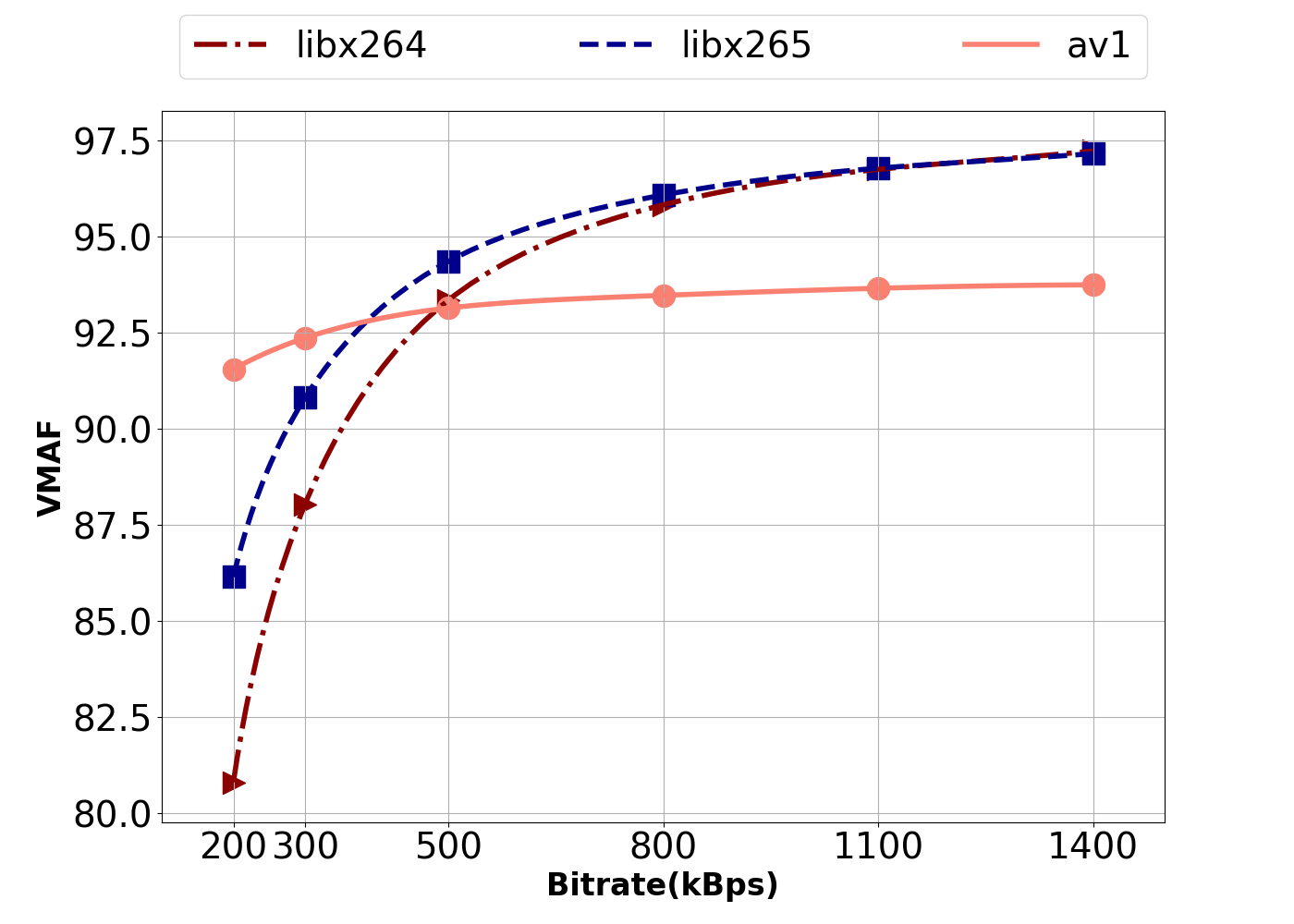}
 \includegraphics[width=0.5\textwidth]{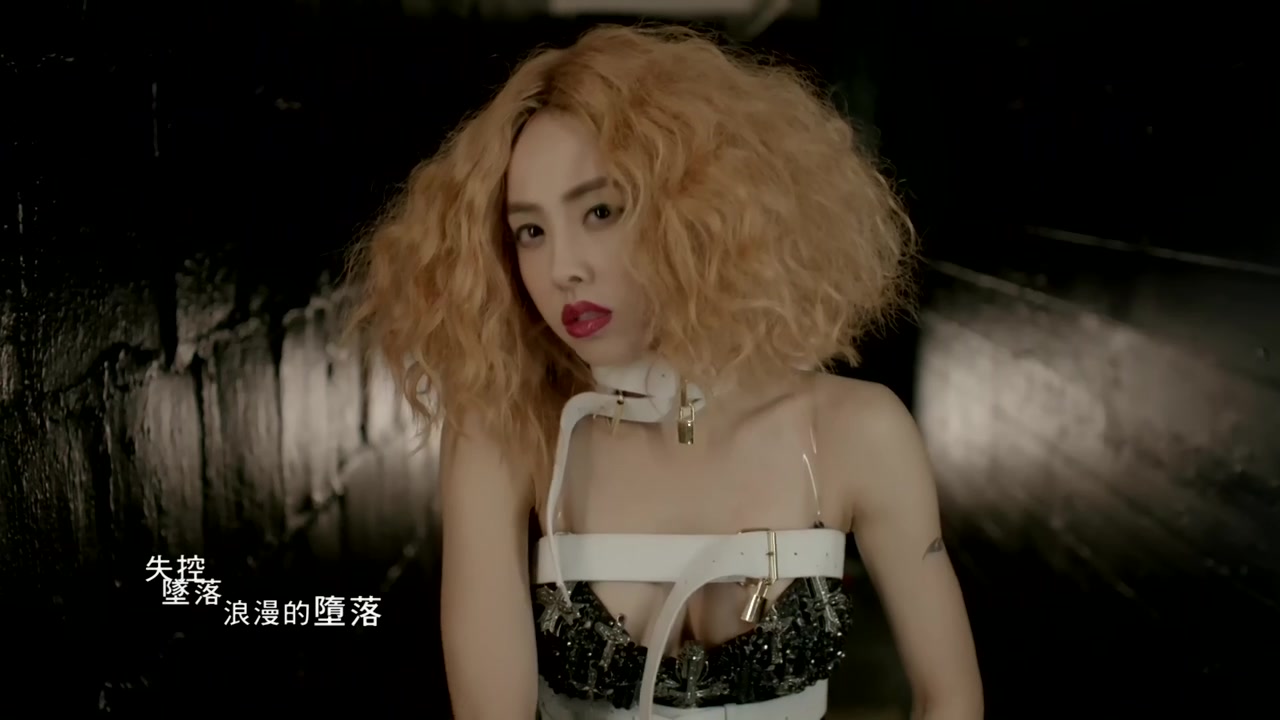}}
  \end{minipage}
  \caption{This group of figures shows our motivation: In the real-time live streaming scenario, high video bitrate is equaled to high video quality, however, in some circumstance, high video quality only requires a low bitrate.}
  \label{fig:vmaf}
\end{figure}

Recent years have witnessed a rapid increase in the requirements of real-time video streaming~\cite{cisco}. Live video streams are being published and watched by different applications(e.g., Twitch, Kwai, Douyu) at any time, from anywhere, and under any network environments. Due to the complicated environment and stochastic property in various network conditions, transmitting video stream with high video bitrate and low latency has become the fundamental challenge in real-time video streaming scenario. Many rate control approaches have been proposed to tackle the problem, such as loss-based approach (TFRC~\cite{handley2002tcp}, RAP~\cite{752152}), delay-based approach (Vegas~\cite{brakmo1995tcp}, LEDBAT (Over UDP)~\cite{rossi2010ledbat}), and model-based approach~(Google Congestion Control(GCC)~\cite{carlucci2016analysis},  Rebera~\cite{kurdoglu2016real}). The same strategy of them is to select bitrate as high as possible with the permission of network condition. However, due to the inequality between high video quality and high bitrate, this strategy may cause a large waste of bandwidth resources. For example, if a video footage consists of darkness and few objects, a low bitrate may also provide a barely satisfactory perceptual video quality but can save large bandwidth resources, and the example is shown in Figure~\ref{fig:vmaf}(a).

In this paper, we propose QARC(video Quality Awareness Rate Control), a novel deep-learning based rate control algorithm aiming to obtain high video quality and low latency. 
Due to that fixed rules fail to effectively handle the complicated scenarios caused by perplexing network conditions and various video content, we leverage DRL-based method to select the future video bitrate, which can adjust itself automatically to the variety of its inputs.
In detail, QARC uses DRL method to train a neural network to select the bitrate for future video frames based on past time network status observed and historical video frames. However, if we directly import raw pictures as the inputs of state, the state space will cause ``state explosion''~\cite{Clarke2012}.
To overcome this, we meticulously divide this complexed RL model into two feasible and useful models:
one is Video Quality Prediction Network (VQPN), which can predict future video quality via previous video frames; the other is Video Quality Reinforcement Learning (VQRL). VQRL uses A3C~\cite{mnih2016asynchronous}, a DRL method, to train the neural network. The inputs of the VQRL are past time network status observed and future video quality predicted by VQPN, and the output is the bitrate for the next video with high video quality and low latency.

We design the training methodologies for those two neural networks respectively. To train VQPN, in addition to some general test video clips, we build up a dataset consisting of various types videos including movie, live-cast show, and music video. For training VQRL, we propose an offline acceleration network simulator to emulate real-world network environment with a trace-driven dataset. We then collect a corpus of network traces for the simulator with both packet-level traces and chunk-level public traces. 

After deciding the architecture of two neural networks,
we compare QARC with existing proposed approaches, results of trace-driven emulation show that QARC outperforms with existing proposed approaches, with improvements in average video quality of 18\% - 25\% and decreases in average queuing delay of 23\% - 45\%.
Besides that, by comparing the performance of QARC with the baseline which represents the offline optimal based on high bitrate and low latency over different network conditions and videos, we find that in all considered scenarios, despite a decrease in average video quality of only 4\% - 9\%, QARC saves the sending rate with 46\% to 60\% and reduces the average queuing delay of 40\% to 50\%.

As a result, our contributions are shown as follows.
\begin{itemize}
\item Unlike the previous goal, we propose a novel sight to evaluate QoE: aiming to optimize video quality rather than video bitrate during the entire video session. 

\item To the best of our knowledge, we are the first to establish a deep reinforcement learning (DRL)  model to select sending bitrate for future video frames based on jointly considered perceptual video quality and network status observed in the real-time video streaming scenario.

\item Due to the complexity of input state, we derive the neural network into two parts: the first part is a neural network used to precisely predict future video quality based on the previous video frames; the second part is an RL model used to determine the proper bitrate based on the output of the first model. By using the output video quality from the first part instead of the raw video frames, the state space of the RL model can be reduced efficiently.

\end{itemize}



\section{Motivation}
In this section, we start by designing an experiment to answer two fundamental questions: 
\begin{itemize}
\item With the enhancement of video encoding technology, what will the correlation change between video quality and video bitrate?
\item Despite the high precision in time series data by using a neural network, can it also precisely predict the fluctuation of the network especially without knowing the saturated bandwidth of the entire video session?
\end{itemize}


\subsection{High Video Quality or High Video Bitrate?}
\label{sec:qualityandbitrate}
To solve this, we establish a testbed to assess the video quality score of selected videos with the given encoding bitrate. The selected videos consist of three video clips, and each of them represents a video with static video scene (live-cast), a video with dynamic video scene (live concert), and a video with hybrid static video scene and dynamic video scene (MV) respectively. 

In our experiment, we use Video Multi-Method Assessment Fusion(VMAF), a smart perceptual video quality assessment algorithm based on support vector machine(SVM)~\cite{rassool2017vmaf}. We compare the video quality score of each video in different encoders. In detail, we use three video encoders in our experiments including x264~\cite{x264}, x265~\cite{x265}, and AV1~\cite{av1}. The first two encoders are popularly used nowadays, and the last one is the state-of-the-art video encoder proposed by Google.

As illustrated in Figure~\ref{fig:vmaf}, comparing VMAF score of different encoders on different videos and encoded bitrates, the results show that as the encode bitrate increases, the rate of increase in video quality score decreases. 
In addition, the refinement of encoder technology does not eliminate this phenomenon. As a result, in the real-time live streaming scenario, if we blindly select the high bitrate, it will make the burden of the network transmission highly increase with little enhancement of video quality.

Inspired by this, we propose a novel sight which aims to optimize perpetual video quality rather than video bitrate during the entire video session.

\begin{figure}[ht]
    \centerline{\includegraphics[width=1.0\linewidth]{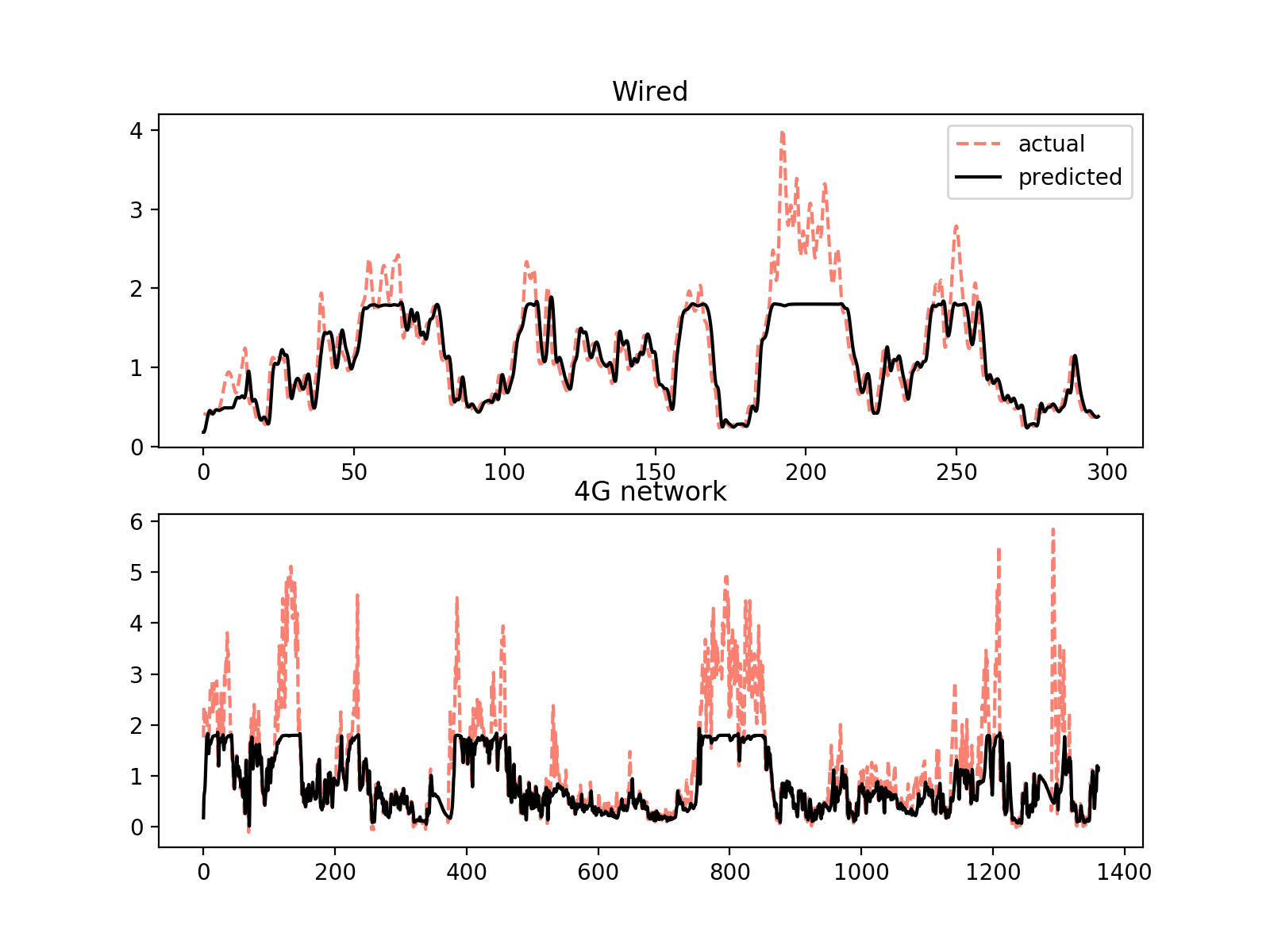}}
    \vspace{-10pt}
    \caption{The Results of Online-learning}
    \label{fig:motivation2} 
\end{figure}

\subsection{Estimating Future Network Status using Neural Network}
The second issue is focused on conventional network congestion control. In the real-time network scenario, by using a neural network, how to accurately estimate the future saturated bandwidth based on past network status observed is still a challenge. In this paper, we use machine learning approach to solve the problem, and in details, we use online learning to train a neural network model to predict the future network status.

Considering past $k$ time-slots, we define $I_t = \{s,r,d\}$ as the input of neural network, where $s$ is the sending rate of past $k$ time-slots measured by the sender; $r$ represents the receiving throughput collected by the receiver of past $k$ time-slots, and $d$ is the delay gradient computed by the receiver at that time-slot. In our experiment, we set $k = 5$. The output is a linear value described as the throughput of next time-slot $t+1 $, and in our problem, this value is equal to the available bandwidth. The model is mainly constructed as a 1D-convolutional network (1D-CNN). To train this model, We propose a network simulator which can use saturated traces to generate network status data, and more details can be seen in Section~\ref{sec:simulator}. In particular, sending rate is constrained in the range of $[0.01,1.8]$ Mbps, which cannot reach the maximum size of available bandwidth.

Figure~\ref{fig:motivation2} illustrates our results on real-world network datasets (Section.~\ref{sec:networkdataset}). As shown, the model that is trained on the synthetic dataset is able to generalize across network conditions, and achieving SMAPE (Eq.~\ref{eq:smape}) score within 11.1\% of the model trained directly on the real-world networks including wired network and 4G network. These results suggest that, in practice, the neural network using 1d-CNN will have an ability to estimate future network status without measuring available bandwidth.


\section{System Architecture}

We start with introducing the conventional end-to-end transmission process for real-time video streaming. The system contains a sender and a receiver, and its transport protocol mainly consists of two channels: the streaming channel and the feedback message channel. At the beginning, the sender deploys a UDP socket channel to send the instant real-time video streaming packets $P = \{p_0,p_1,\cdots,p_k\}$, denoted as a packet train~\cite{sato2017experimental}, to the receiver through the streaming channel. The receiver then feeds network status observed back to the sender through the feedback channel. Based on this information, the sender will select bitrate for next time period. 
\begin{figure}
    \centerline{\includegraphics[width=1.1\linewidth]{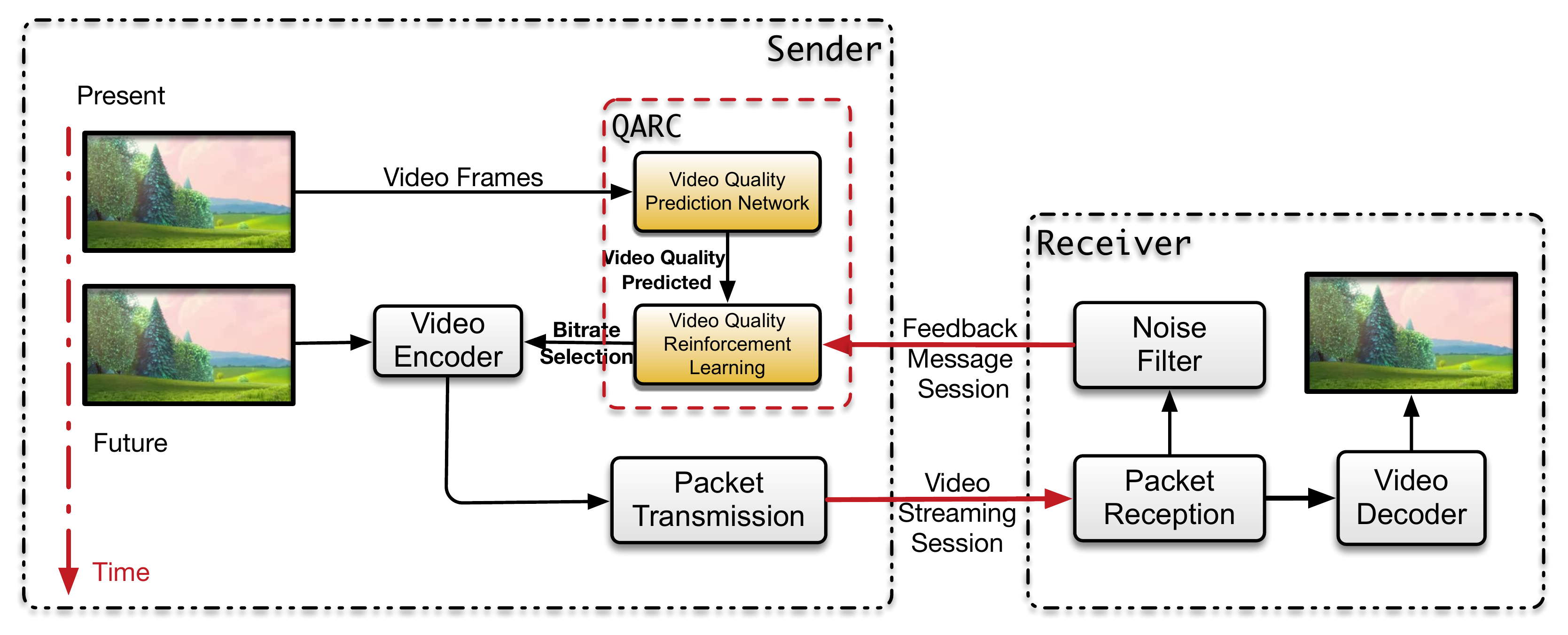}}
    \caption{QARC's System Architecture}
    \label{fig:overview} 
\end{figure}

As shown in Figure~\ref{fig:overview}, on the basis of conventional real-time video streaming system architecture, we propose QARC, which is placed on the sender side. Motivated by the unbalanced growth of video quality and video bitrate as described in Section~\ref{sec:qualityandbitrate}, we design a RL model to ``learn'' the correlation among the previous video frame, network status, and the best future bitrate. However, if we use raw pictures directly as its inputs, the state will cause ``state explosion''~\cite{Clarke2012}. Moreover, it will hard to train and validate in an allowable time. To overcome this, we meticulously divide the complexed RL model into two feasible and useful models, which involves:


\textbf{Video Quality Prediction Network(VQPN)}, proposed by end-to-end deep learning method, which predicts the future video quality metrics based on historical video frames;

\textbf{Video Quality Reinforcement Learning(VQRL)}, which uses A3C, an effective actor-critic method which trains two neural networks to select bitrates for future video frames based on network status observations and the future video quality metrics predicted by VQPN.

\begin{figure}[ht]
    \centerline{\includegraphics[width=1.0\linewidth]{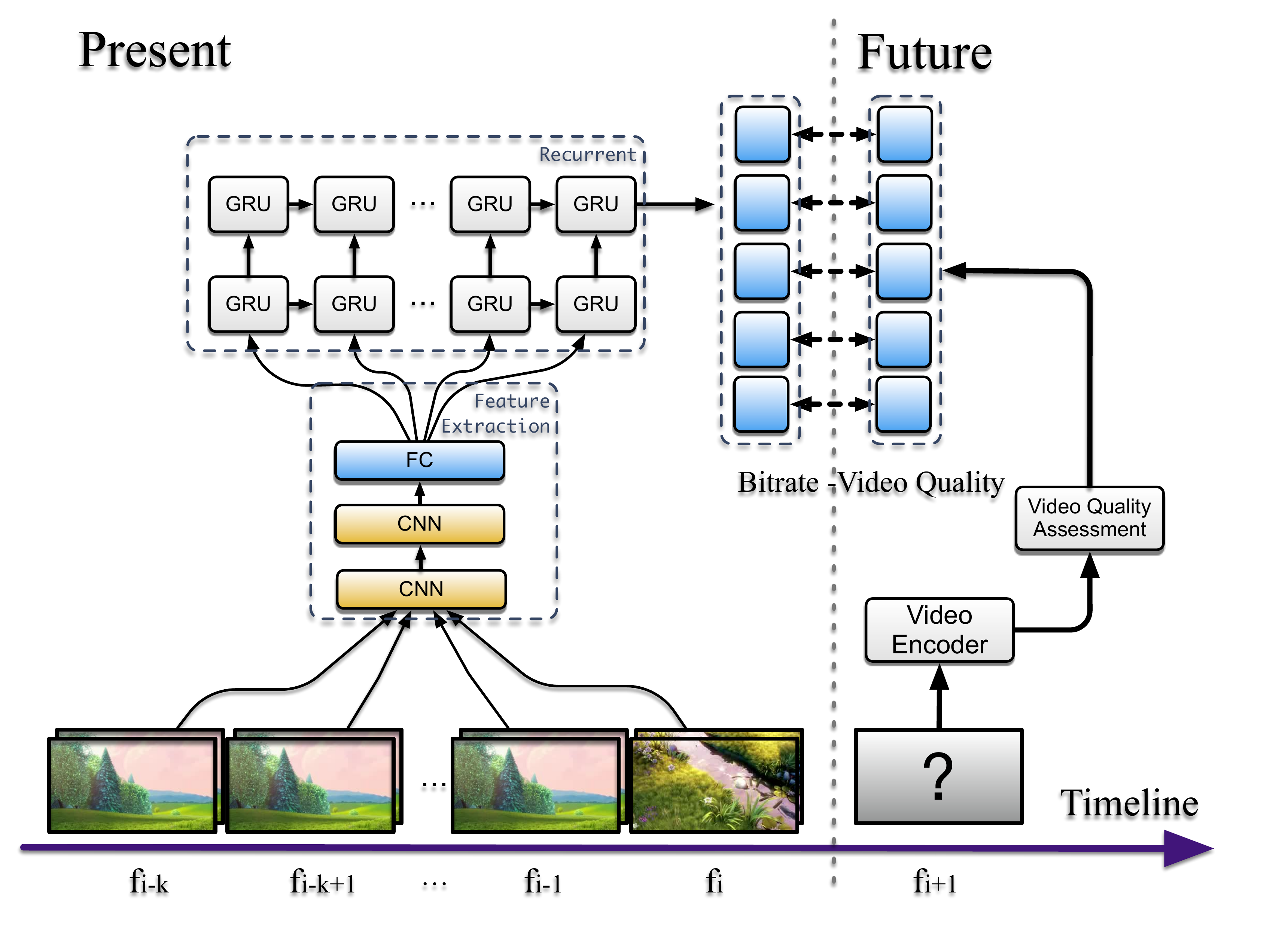}}
    \caption{VQPN Architecture Overview}
    \label{fig:VQP}
\end{figure}
\subsection{Video Quality Prediction Network(VQPN)}

To help the RL model select a proper encoding bitrate for the next frame, we need to let the model ``know'' the relationship between the bitrate and corresponding video quality first. 
However, this form of prediction is quite challenging, because the perceptual video quality is closely related to the video itself. As shown in Figure~\ref{fig:vmaf}, the video type, brightness, and objects number all have a great impact on the correlation between bitrate and VMAF.
Motivated by the effectiveness of the neural network in a prediction of time sequence data, we design video quality prediction network(VQPN) helps the RL model to predict the perceptual video quality of the future frame. Figure~\ref{fig:VQP} describes the VQPN's neural network architecture, which is mainly made up with a layer that extracts image features through Convolutional Neural Network (CNN), and another layer which capture temporarily features via Recurrent Neural Network (RNN). Details are shown as follows.



\textbf{Video Quality Metric:}
We use mean video quality metric to describe the quality of the video over a period. For each raw video frame $f_i$ in time-slot t, the video quality score $V_{f_i,bitrate}$ is computed by the raw video frames $f$ and the bitrate at which the raw video frames $f$ will be encoded, then the mean score $V_{t,bitrate}$ is defined as the average value of $V_{f,bitrate}$. In our study, we use mean VMAF score, which is a score that is specifically formulated by Netflix to correlate strongly with subjective MOS scores to describe the video quality of video frames. In particular, we normalize the score into the distribution of the range from [0,1].


\textbf{Input.}~VQPN takes state inputs $F_i = [f_{i-k}, f_{i-k+1},\cdots,$ $f_{i}]$ to its neural network, in which $f_i$ reflects the i-th sampled video frame.


\textbf{Extract image features:} VQPN uses CNN layers to extract frame features, which can obtain the spatial information for each video frame in inputs $F_i$.

\textbf{Capture temporal features:} Upon extracting frame features, VQPN uses a double-layered recurrent layer~\cite{chung2014empirical} to further extract temporal characteristics of the video frames $F_i$ in past k sequences.

\textbf{Output:} The outputs of VQPN are the prediction of the video quality assessment in the next time slot $t+1$ of candidate bitrates, denoted as $V_{t+1}$.

\textbf{Loss function:} We use mean square error(MSE) to describe the loss function, besides that, we also consider to add regulation to the loss function to decrease the probability of over fitting that on training set. Let $\hat{V_{t}}$ denote the real vector of video quality score of the video in time {t}. Therefore, the loss function can be written as (Eq.~\ref{eq:loss}), where $\lambda$ is the regulation coefficient.

\begin{align}
L_t(V;\theta) = \frac{1}{N}\sum|V_{t} - \hat{V_{t}}|^{2} + \lambda||\theta||^{2}
\label{eq:loss}
\end{align}

\subsection{Video Quality Reinforcement Learning(VQRL)}

In our study, we aim to let the neural network ``learn'' a video bitrate selection policy from observations instead of using preset rules in the form of fine-tuned heuristics. Specifically, our approach is based on RL. The sender, serving as an agent in RL problem, observes a set of metrics including future video quality and previous network status as the state. The neural network then selects the action as the output which denotes the video bitrate of next time-slot. Then the goal is to find the policy that maximizes the quality of experience (QoE) perceived by the user. In our scheme, QoE is influenced by video quality, latency, and smoothness. 

As shown in Figure~\ref{fig:VQRL-INTRO}, We formulate ``video quality first'' real-time video streaming problem within A3C framework, named as video quality reinforcement learning (VQRL). Detailing our system components, which include:

\begin{figure}
 \centerline{\includegraphics[width=0.7\linewidth]{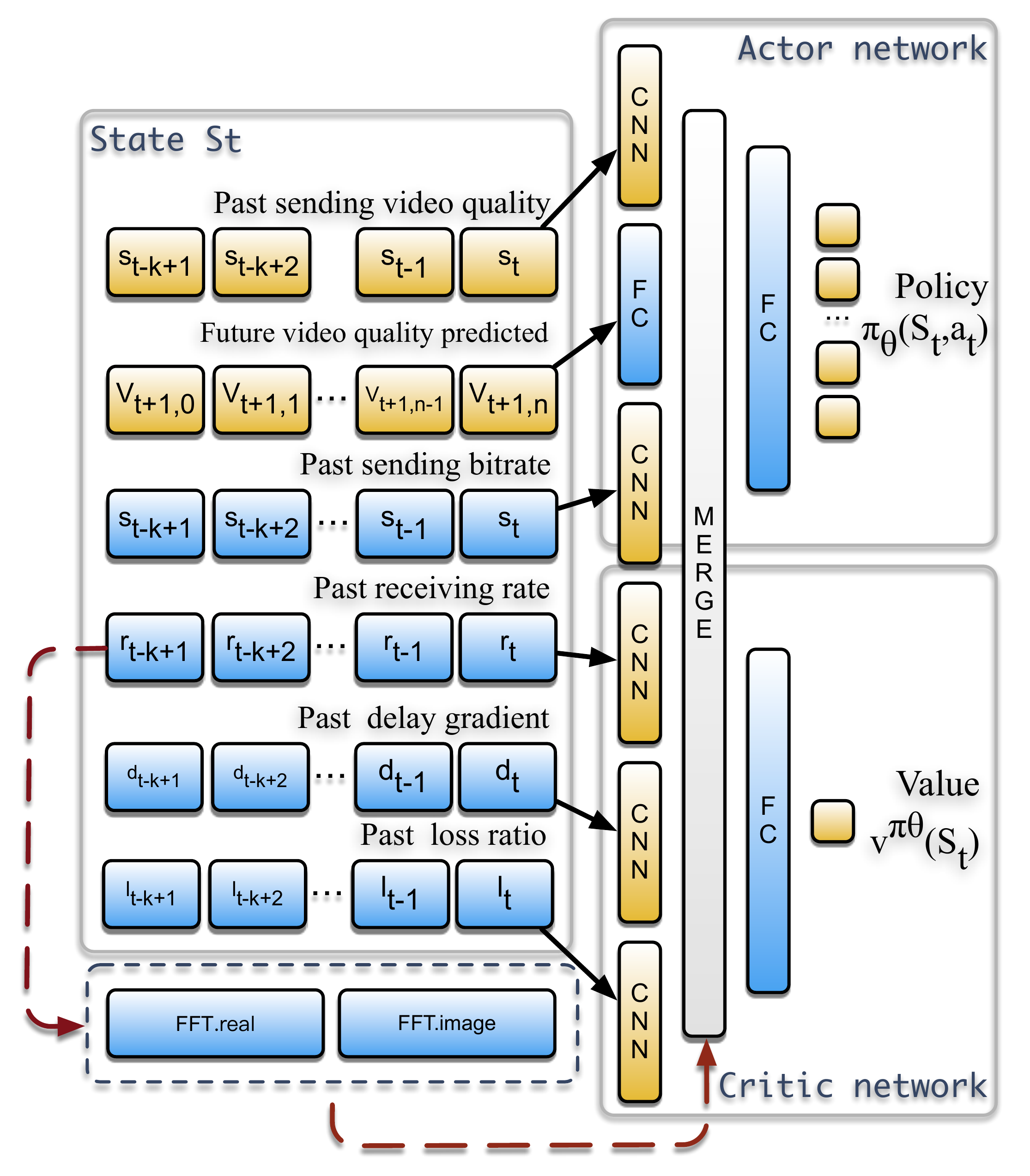}}
    \caption{The Actor-Critic algorithm that VQRL uses to generate sending bitrate selection policies}
    \label{fig:VQRL-INTRO} 
    \vspace{-15pt}
\end{figure}

\textbf{State:} We consider the metrics which can be obtained by both sender and receiver from feedback message session. VQRL's learning agent pushes the input state of time-slot $t$ $s_t = \{p,v,s,r,d,l\}$ into neural network, where $p$ means the past sending video quality, $v$ represents the future video quality predicted by VQPN, $s$ is the video sending rate of the past k sequences which is equal to the throughput measurement from the uplink of the sender; $r$ represents the receiving bitrate of past $k$ sequences measured by the receiver; $d$ is the delay gradient which is measured between a sender and receiver of the recent $k$ sequences; $l$ is the packet loss ratio of the previous $k$ sequences. 

To better estimate the network condition in our scenario, we need precisely measure queuing delay of each packet. However, due to the clocks on both sides
are unsynchronized, the measurements are unreliable. Motivated by \cite{carlucci2016analysis}, we also use delay gradient to solve the problem. More details can be seen in \citep{carlucci2016analysis,HuangZZS18}.



Besides that, we assume receiving bitrate as a form of signal. Then, the Fast Fourier Transform (FFT) can be used to decompose signals into a complex-valued function of frequency, whose absolute value represents the amount of that frequency present in the original function, whose complex argument is the phase offset of the basic sinusoid in that frequency.~\cite{frigo1999a} 
As a result, we add the additional features into input which decomposed the receive rate sequence through FFT. The results that validate its improvement will be discussed in Section~\ref{sec:VQRL_exp};

%
%
\textbf{Action:} The agent needs to take action when receiving the state, and the policy is the guide telling the agent which action will be selected in the RL problem. In general, the action space is discrete, and the output of the policy network is defined as a probability distribution: $f(s_t, a_t)$, meaning the probability of selection action $a_t$ being in state $s_t$. In this paper, the action space contains the candidate of sending bitrate in the next time-slot $t$. 

In traditional RL problem, the state space is small and can be represented in a tabular form, and there have been a lot of effective algorithms to solve this kind of problems, such as Q-learning and SARSA~\cite{sutton1998reinforcement}. However, in our problem, the state space is fairly large, e.g., loss rate and received bit rate are continuous numbers, so it is impossible to store the state in a tabular form. To tackle this barrier, we use a neural network~\cite{hagan1996neural} to represent the policy, and the weights of the neural network, we use $\theta$ in this paper, are called the policy parameters. In recent researches, the technique of combining neural network and RL is widely used to solve large-state-space RL problems~\citep{silver2016mastering,mao2017neural} and shows its exceptional power;

\textbf{Reward:}~Our reward~(QoE) will be described in Section~\ref{sec:qoe};

\textbf{Training:} In the RL problem, after taking a specific action in state $s_t$, the agent will get a corresponding reward , and the goal for the RL agent is to find the best action in each state which can maximize the accumulated reward $r_t$ and as a result, the policy should be changed in the direction of achieving this goal. In this paper, we use A3C~\cite{mnih2016asynchronous}, a state of the art actor-critic RL algorithm, as the fundamental algorithm of our system, and in this algorithm, policy training is done by performing policy gradient algorithm.

The key thought of the policy gradient algorithm is to change the parameter in the direction of increasing the accumulated reward. The gradient direction is the direction in which a function increases. The gradient of the accumulated reward with respect to policy parameter $\theta$ can be written as:

\begin{align}
\nabla E_{\pi_{\theta}}[\sum_{t=0}^{\infty} \gamma^t r_t ]=E_{\pi_{\theta}}[\nabla_{\theta}log_{\pi_{\theta}}(s,a)A^{\pi_{\theta}}(s,a)]  
\end{align}

We can use:~$E_\theta[\nabla_\theta log{\pi_\theta(s,a)}A^{\pi_\theta}(s,a)]$ as its unbiased form, where $A(s_t,a_t)$is called the advantage of action $a_t$ in state $s_t$ which satisfies the following equality: 

\begin{align}
A(a_t, s_t)=Q(a_t,s_t)-V(s_t)
\end{align}

Where $V(s_t)$ is the estimate of the value function of state $s_t$ and $Q(a_t, s_t)$ is the value of taking certain action at in state $s_t$, and it can also be written as:

\begin{align}
Q(a_t,s_t)=r_t+\gamma V(s_{t+1}|\theta_ {t+1})
\end{align}

\noindent Thus, policy parameter will be updated as:

\begin{align}
\theta \gets \theta + \alpha \sum_{t}\nabla_\theta log\pi_{\theta}(s_t,a_t)A(s_t,a_t)
\end{align}

\noindent in which the parameter $\alpha$ represents the learning rate. To calculate $A(s_t, a_t)$, we need to have the $V(s_t)$ first, and we can estimate it in the value network.
The value network aims to give a reasonable estimate of the actual value of the expected accumulated reward of state $s_t$, written as $V(s_t|\theta_v)$. Continuing the same line of thought, value network also uses neural network to represent the large state space. In this paper, we use n-step Q-learning to update the network parameter \cite{mnih2016asynchronous}, and for each time, the error between estimation and true value can be represented as $Err_t=(r_t+\gamma V(s_{t+1}|\theta_v)-V(s_t|\theta_v))^2$, where $V(s_t|\theta_v)$ is the estimate of $V(s_t)$, and to reduce the $Err_t$, the direction of changing parameter $\theta_v$ is the negative gradient of it, and in A3C, the gradient will be added up with respect to $t$, so the value network will be updated as:

\begin{align}
\theta_v \gets \theta_v - \sum_t \nabla_{\theta_v} Err_t  
\end{align}

\noindent where $\alpha$ is the learning rate. Inspired by~\cite{mao2017neural,mnih2016asynchronous}, we also add the entropy of policy in the object of policy network, which can effectively discourage converging to suboptimal policies. See more details in~\cite{mnih2016asynchronous}. So the update of $\theta$ will be rewritten as:

\begin{align}
\theta \gets \theta + \alpha \sum_t \nabla log_{\pi_\theta}(s_t,a_t)A(s_t,a_t)+\beta \nabla_\theta H(\pi_\theta(\cdot|s_t)) 
\end{align}

\noindent where $\beta$ is also a hyper-parameter, $H(\cdot)$ is the entropy of the policy. After convergence, the value network will be abandoned, and we only use policy network to make decisions;

\textbf{Multiple training:} To accelerate the training process, as suggested by \cite{mnih2016asynchronous}, we modify VQRL's training in the single agent as training in multi-agents. Multi-agents training consists of two parts, a central agent and a group of forwarding propagation agents. The forward propagation agents only decide with both policy and critic via state inputs and neural network model received by the central agent for each step, then it sends the $n$-dim vector containing $\{state, action, reward\}$ to the central agent. The central agent uses the actor-critic algorithm to compute gradient and then updates its neural network model. Finally, the central agent pushes the newest model to each forward propagation agent. Note that this can happen asynchronously among all agents, for instance, there is no locking between agents. By default, VQRL with multiple training uses 8 forward propagation agents and 1 central agent;


\textbf{Train with network simulator:}
\label{sec:simulator} To train VQRL, we first consider to train our neural network model in real-world network conditions, e.g., deploying the model on the edge server. With the increasing number of session, the model will finally converge. However, training the model online is hard to converge because RL training should meet almost all network status as the state. We then decide to train the model in simulated offline networks. Hence, we are a facing new challenge: How to design a fast-forward network simulator which can precisely compute the latency with given saturated trace and sending rate?


\begin{figure}
    \centerline{\includegraphics[scale=0.4]{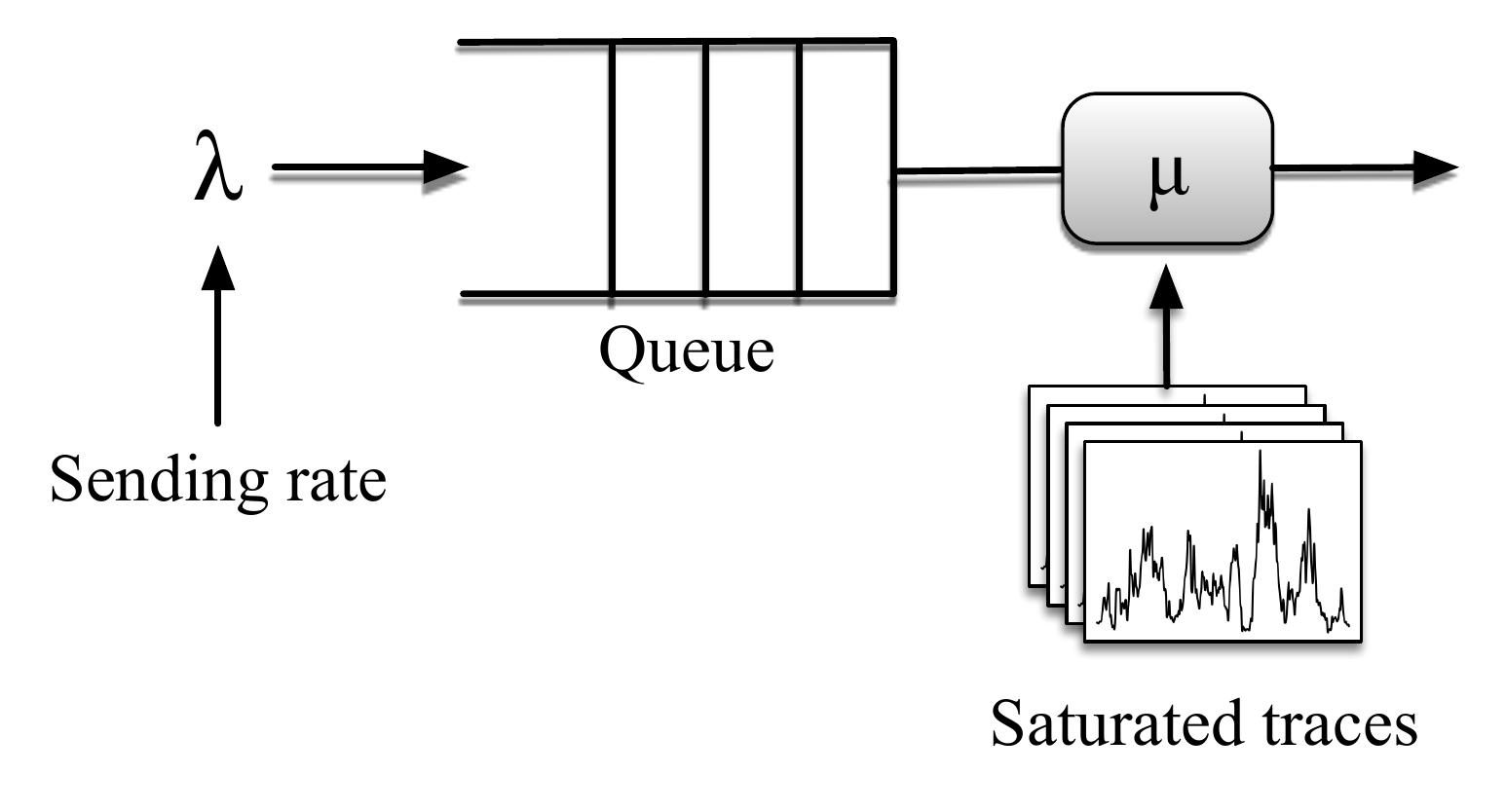}}
    \caption{The working principle of the network simulator.}
    \label{fig:queue} 
    \vspace{-20pt}
\end{figure}

To train our model, our training data should consist of queuing delay rather than one-way delay. So, our simulator should simulate the process of the packets coming and leaving in different network conditions, and keep track of the timestamps, by which we can get the corresponding queuing delay.   
Inspired by ~\cite{winstein2013stochastic} and ~\cite{netravali2015mahimahi:}, we use saturated network trace to generate queuing delay data. Seen in Figure~\ref{fig:queue}, assuming the distribution of packets arrival and leave fits closely to the Poisson process~\cite{winstein2013stochastic}, we use sending bitrate and bandwidth in saturated network traces as the arriving rate $\lambda$ and leaving rate $\mu$, respectively.

\begin{figure*}
  \centering
  \begin{minipage}{0.33\linewidth}
      \centering
      \subfigure[The curves of average reward under different neural network model including CNN, FNN, and GRU.]{\includegraphics[width=0.9\textwidth]{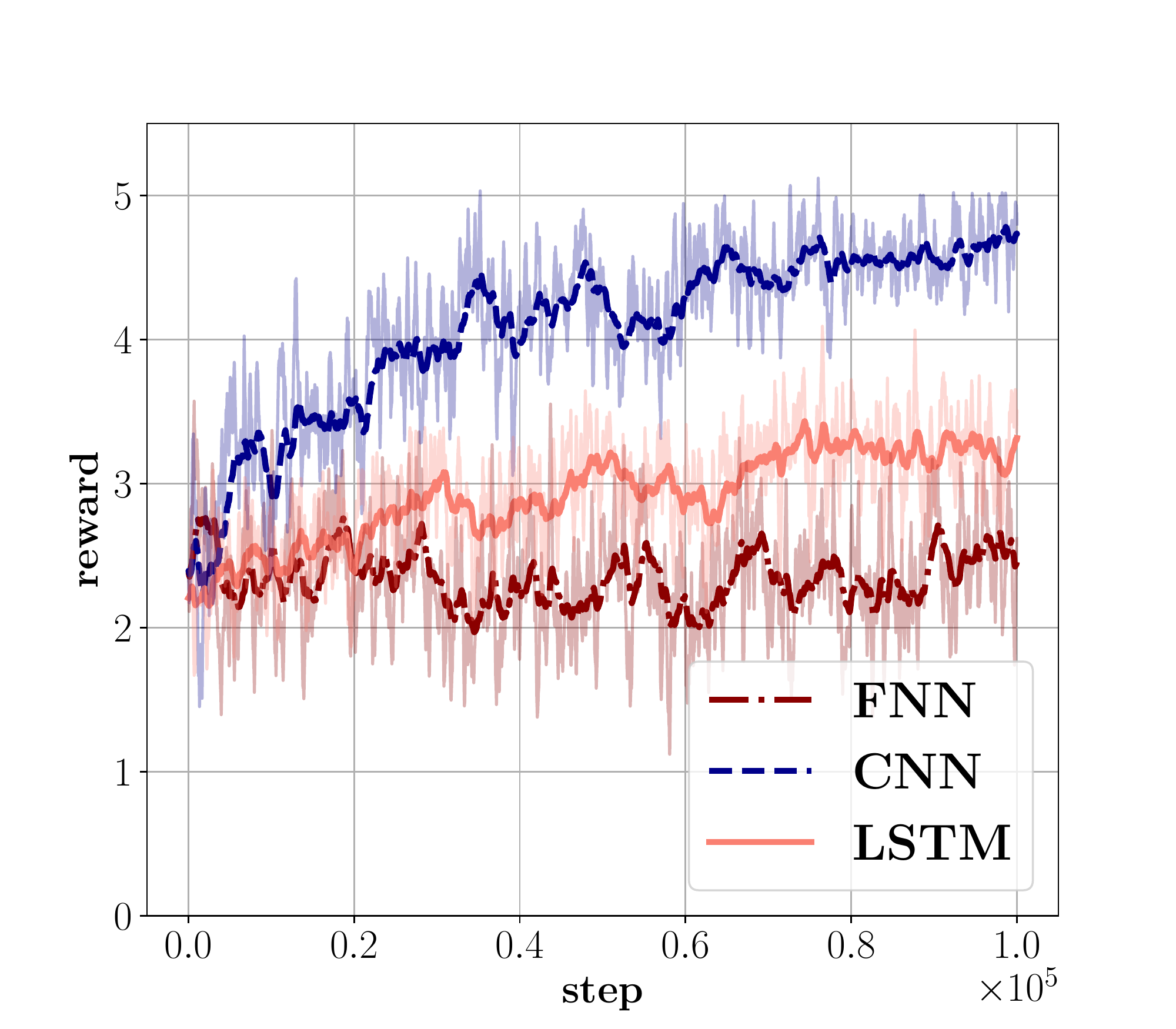}}
  \end{minipage}
  \begin{minipage}{0.33\linewidth}
      \centering
      \subfigure[Comparing VQRL which uses FFT with the one without using it.]{\includegraphics[width=0.9\textwidth]{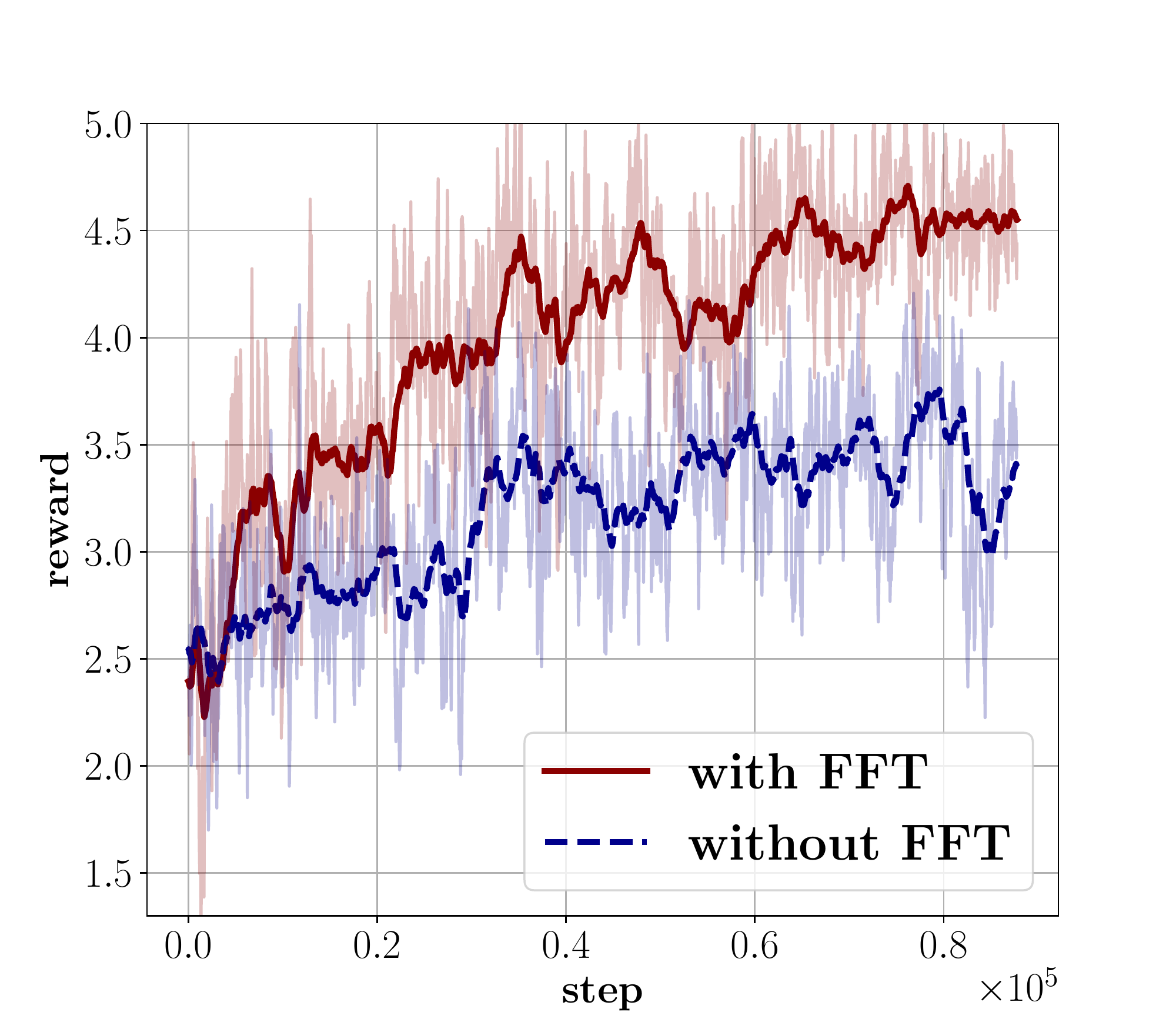}}
  \end{minipage}  
  \begin{minipage}{0.33\linewidth}
      \centering
      \subfigure[Sweeping sequence length and number of filters in VQRL's neural network architecture.]{\includegraphics[width=0.9\textwidth]{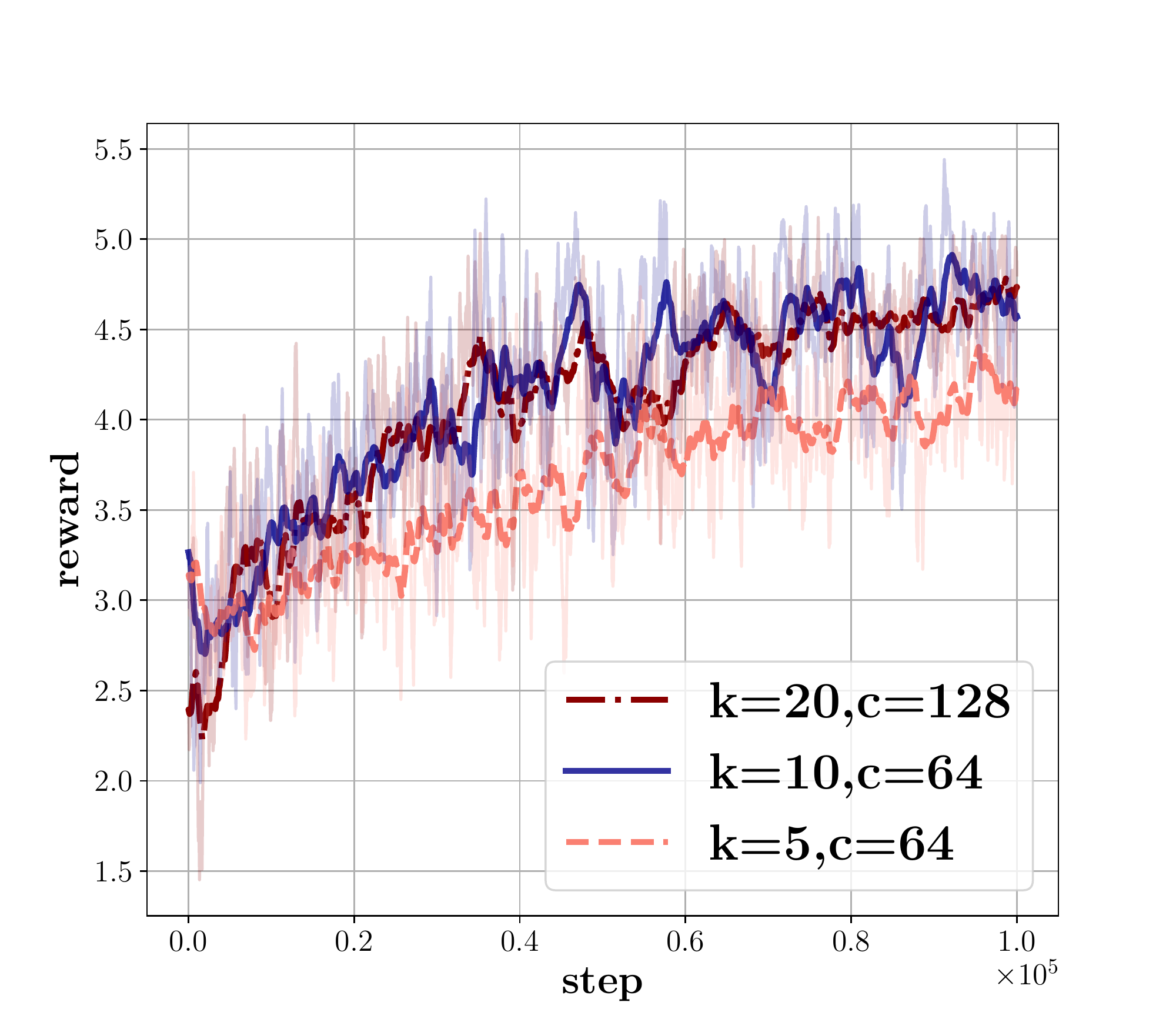}}
  \end{minipage}
  \caption{VQRL's implementation}
  \label{fig:VQRL}
\end{figure*}

\section{Evaluation}

\subsection{Datasets and Metrics}

\textbf{Video dataset:}
We train and test VQPN on two video datasets, that is, VideoSet: a large-scale compressed video quality dataset based on JND measurement and self-collected video datasets: a video quality dataset involves live-casts, music-videos, and some short movies. For each video in datasets, we measure its VMAF with the bitrate of 300Kbps to 1400Kbps, and the reference resolution is configured as $800\times480$, which is the same size as default resolution that observed by the receiver during the real-time live streaming. We generate the VMAF video datasets using both x264 and x265 encoder. 

\textbf{Network traces:} 
\label{sec:networkdataset}
To train and evaluate VQRL, the first thing we must do is to generate saturated network trace datasets. However, these types of network traces are hard to be recorded, even public datasets are extremely limited. For example, Cellsim~\cite{winstein2013stochastic} only provides a small number of saturated network traces which describe the cellular network conditions instead of all network environments, which hardly afford us to make our neural network converge. Thus, we consider to collect datasets in two ways: 

\begin{itemize}
\item \textbf{Packet-level network traces:} We use a proprietary dataset of packet-level live-cast session status from all platforms APPs of Kwai collected in January 2018. \footnote{Kwai is a leading platform in China which has over 700 million users worldwide, and millions of original videos are published on it every day. } The dataset, recorded as packet train, consists of over 14 million sessions from 47,000 users covering 50 thousand unique sessions over three days in January 2018. For each session, it is consists of packet size, packet send time and packet receive time. Based on raw data collected, we propose measuring the available bandwidth {ABW/n} of the whole link, where the available sample bandwidth is obtained by the packet train which is received in the receiving side in period n. Motivated by the one-way-delay estimation method in Ledbat~\cite{rossi2010ledbat}, We generate 2,300 real network traces from packet train datasets.
%
%
%
%
\item \textbf{Chunk-level network traces:} We also collect hybrid network traces datasets which consists of different network datasets, such as FCC~\cite{bworld} and Norway~\cite{riiser2013commute}. The FCC dataset is a broadband dataset, and Norway dataset is mainly collected in 3G/HSDPA environment. In short, we generate 1,000 network traces from the datasets. 
\item \textbf{Synthetic network traces:}We generate a synthetic dataset using a Markovian model where each state represented an average throughput in the aforementioned range.\cite{mao2017neural} Thus, we create a dataset in over 500 traces which can cover a board set of network conditions.

\end{itemize}

\textbf{QoE metrics:} For a better result, we consider designing Quality of Experience (QoE) metric based on previous scheme. In the recent research~\cite{mao2017neural}, QoE metrics are evaluated as a method with 4 essential factors: bitrate received, loss ratio, latency, and delay gradient, without considering video quality metric. Still, in this paper, after rethinking the correspondence between video quality and video bitrate,  we redefine the QoE metric as (Eq.~\ref{eq:qoe})
\label{sec:qoe}
\begin{align}
\texttt{QoE} = \sum_{n=1}^{N}{(V_n -\alpha B_n - \beta D_n)} - \gamma \sum_{n=1}^{N-1}{|V_n - V_{n-1}|}
\label{eq:qoe}
\end{align}

\noindent for a live video with N time-slots. Where $V_n$ denotes the video quality of time $n$, $B_n$ is the video bitrate that the sender selects, and $D_n$ represents the delay gradient measured by the receiver. The final term comprises the smoothness of video quality. Coefficient $\alpha, \beta$ and $\gamma$ are the weight to describe their aggressiveness.

\begin{table}
\begin{center}
\begin{tabular}{cc|p{0.8cm}p{0.8cm}p{0.8cm}p{0.8cm}}
\toprule
\multirow{2}{*}{\textbf{filter number}} &
    \multirow{2}{*}{\textbf{hidden units}}  &
    \multicolumn{4}{ c }{\textbf{Learning Rate}} \\
    \cline{3-6} 
    & & 1e-3 & \textbf{1e-4} & 1e-5 & 6e-6\\
\midrule
32 & 32 & 4.88 & 5.20 & 4.42 & 4.24 \\
32 & 128 & 4.40 & 4.28 & 4.24 & 4.13 \\
64 & 64 & 3.94 & 3.93 & 4.22 & 4.31 \\
64 & 128 & 4.92 & 4.17 & 4.16 & 4.17 \\
\textbf{128} & \textbf{64} & 4.20 & \textbf{3.80} & 4.17 & 4.23 \\
128 & 128 & 4.52 & 3.86 & 4.15 & 3.99 \\
\bottomrule
\end{tabular}
\end{center}
\caption{Comparing performance (SMAPE\%) of VQPN with different filter number and hidden units. Results are collected under learning rate=1e-3,1e-4,1e-5, and 6e-6 respectively.}
\label{table:vqpn}
\vspace{-30pt}
\end{table}

\subsection{Implementation}
We now describe the implementation of QARC.
In this section, we decide the best hyper-parameters and explain the implementation of VQPN and VQRL respectively.

\textbf{Time-slot t:} In this paper, we set time-slot $t$ as 1s.



\textbf{VQPN:} The introduced VQPN help VQRL predict future video quality, but we have yet studied how to set the hyper-parameters. Table~\ref{table:vqpn} shows our results with different settings of filter number, hidden units, and learning rate. Results are summarized as symmetric mean absolute percentage error (SMAPE) metric, which is computed as Eq.~\ref{eq:smape}:
 \begin{align}
 \begin{aligned}
    {\text{SMAPE}}={\frac {100\%}{n}}\sum _{t=1}^{n}{\frac {\left|F_{t}-A_{t}\right|}{(|A_{t}|+|F_{t}|)/2}}.
    \label{eq:smape}
 \end{aligned}
 \end{align}

Here $A_t$ is the actual value and $F_t$ is the forecast value. Empirically, filter number = 128, hidden units = 64, and learning rate = 1e-4 yields the best performance.

To sum up, VQPN passes $t=5$ past time video, and it samples 5 frames for each time, totally $k=25$ previous frames as input to the neural network architecture, and each size of the frame is defined as [64,36] with 3 channels. The input frames then extract features in 128-dimension vector via a feature extraction layer respectively. The feature extraction layer is constructed with 5 layers, a conv layer with 64 filters, each of size 5 with stride 1, an average pooling layer with filter number $3\times3$, an another conv layer with 64 filters, each of size 3 with stride 1, also, a max pooling layer with filter number $2\times2$. Finally, the feature extraction layer passes the features into a hidden layer with 64 neurons. 

Considering the frame sequence as a time series data, a recurrent network is designed to estimate future video quality. VQPN passes $k$ = 25 feature maps to a gated recurrent unit(GRU) layer with 64 hidden units, then the states of that layer are passed to another GRU layer with the same hidden units. A hidden layer is then connected to the hidden output of the last GRU layer. Finally, VQPN uses the final output as a 5-dimension vector, and for each value in the vector represents the video quality score of video bitrate $\{300, 500, 800, 1100, 1400\}$ Kbps. 

During the training process, we use Adam gradient optimizer to optimize VQPN with learning rate $\alpha$ = $10^{-4}$. In this work, we use TensorFlow~\cite{abadi2016tensorflow} to implement this architecture, in particular, we leveraged the TFLearn deep learning library's TensorFlow API to declare VQPN.

\textbf{VQRL:} In this section, we describe how to choose the best neural network model of VQRL. Firstly, we design three different models which are based on FNN (Feedback Neural Network), CNN, and LSTM (Long-Short Term Memory) respectively. We set sequence length $k = 5$,We use the QoE metric with $\alpha = 0.2$, $\beta = 1.0$ and $\gamma = 1.0$ as the baseline reward. As illustrated in Figure~\ref{fig:VQRL}(a), the CNN model increase the average QoE by about 39\% compared with the LSTM model and about 83\% compared with the FNN model. 

\begin{figure*}
  \centering
  \begin{minipage}{0.33\linewidth}
      \centering
      \includegraphics[width=1.0\textwidth]{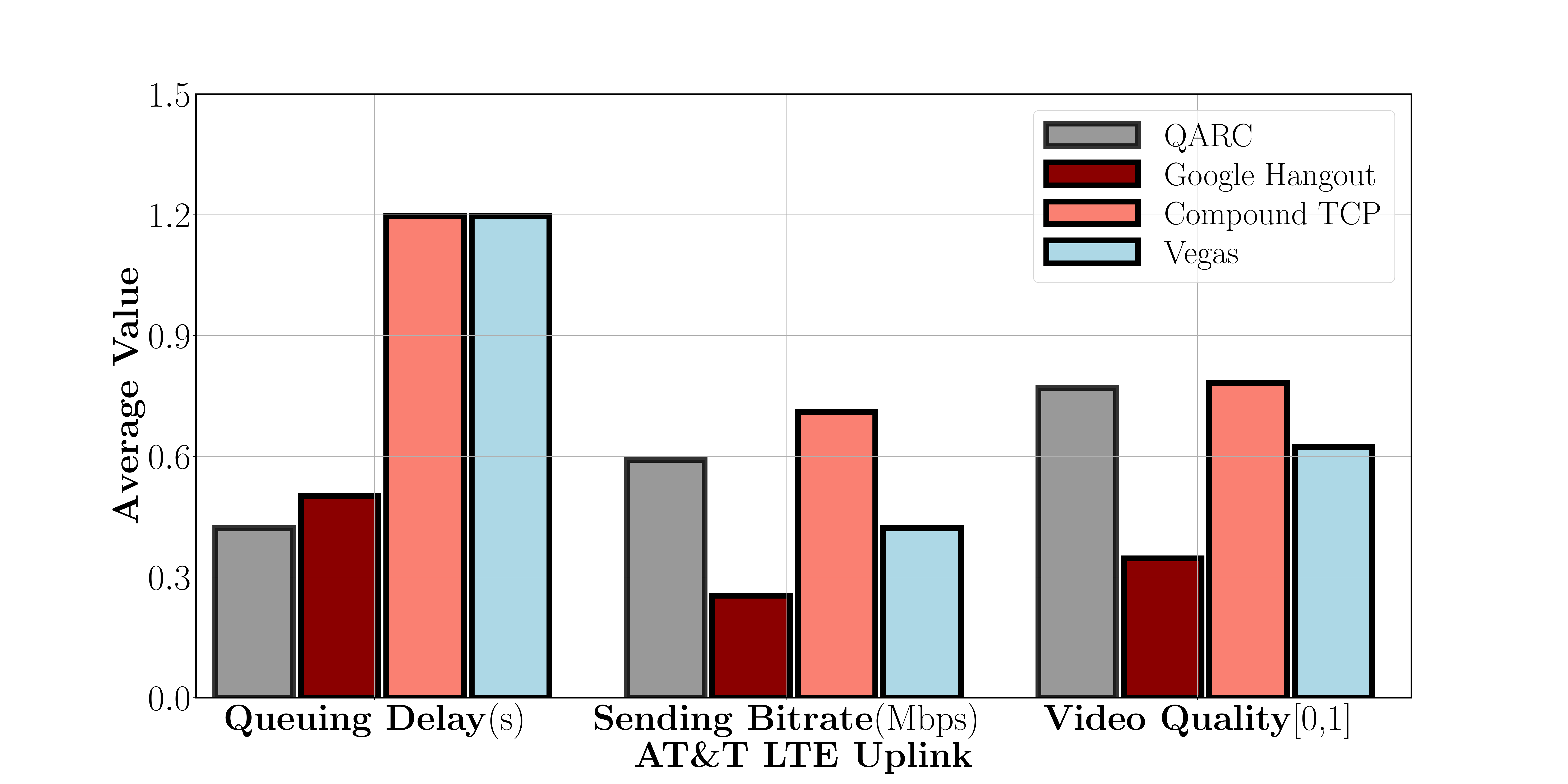}
  \end{minipage}
  \begin{minipage}{0.33\linewidth}
      \centering
      \includegraphics[width=1.0\textwidth]{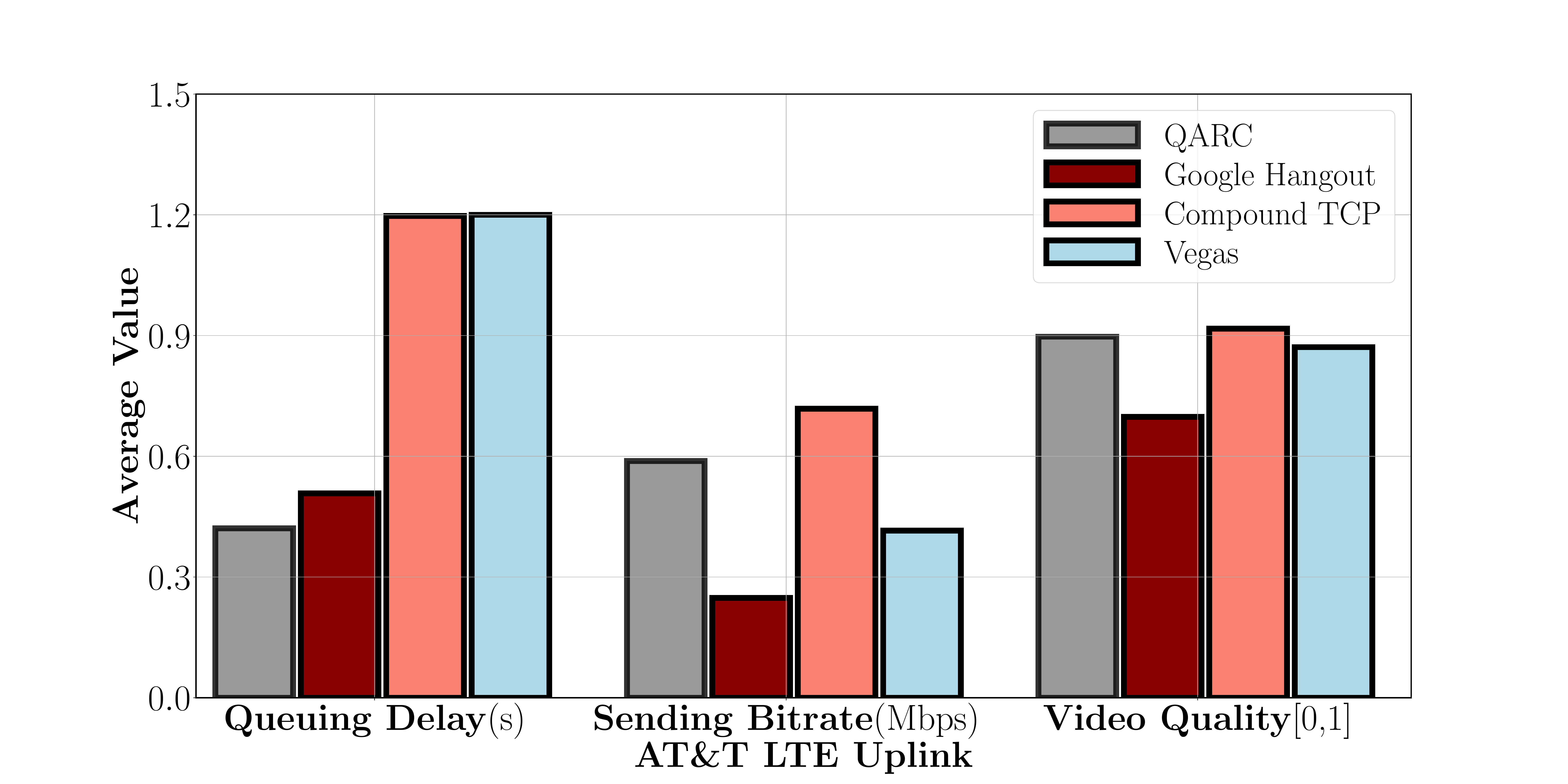}
  \end{minipage}  
  \begin{minipage}{0.33\linewidth}
      \centering
      \includegraphics[width=1.0\textwidth]{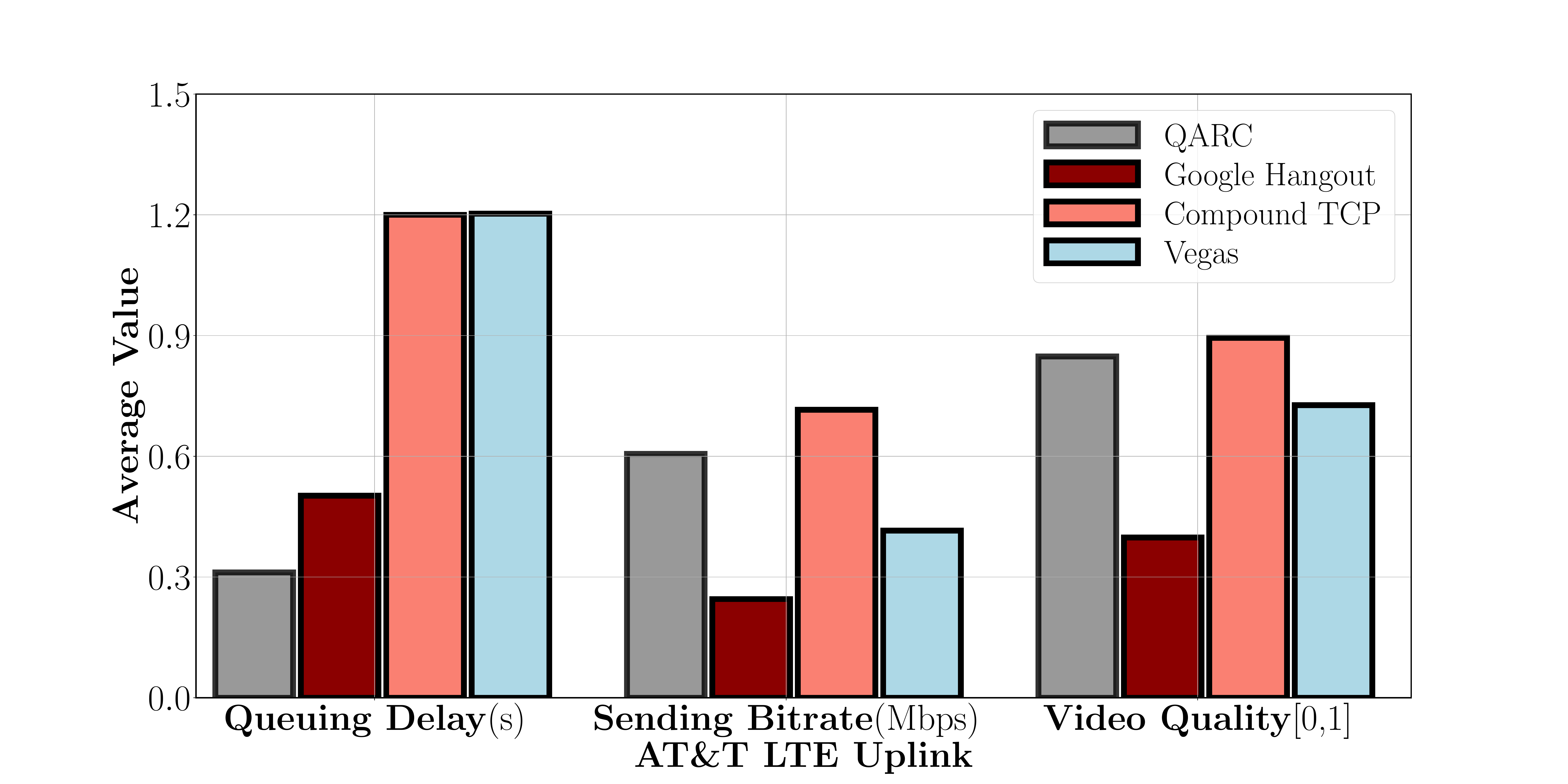}
  \end{minipage}
  \vspace{-10pt}
  \caption{Comparing QARC with previously proposed approaches on the 4G network environments: The QoE of QARC is considered as $\alpha=0.2$, $\beta=10.0$,and $\gamma=1.0$. After testing three video clips, results are shown as average queuing delay, average sending rates, and average video quality.}
  \label{fig:exp3}
  \centering
  \begin{minipage}{0.33\linewidth}
      \centering
      \includegraphics[width=1.0\textwidth]{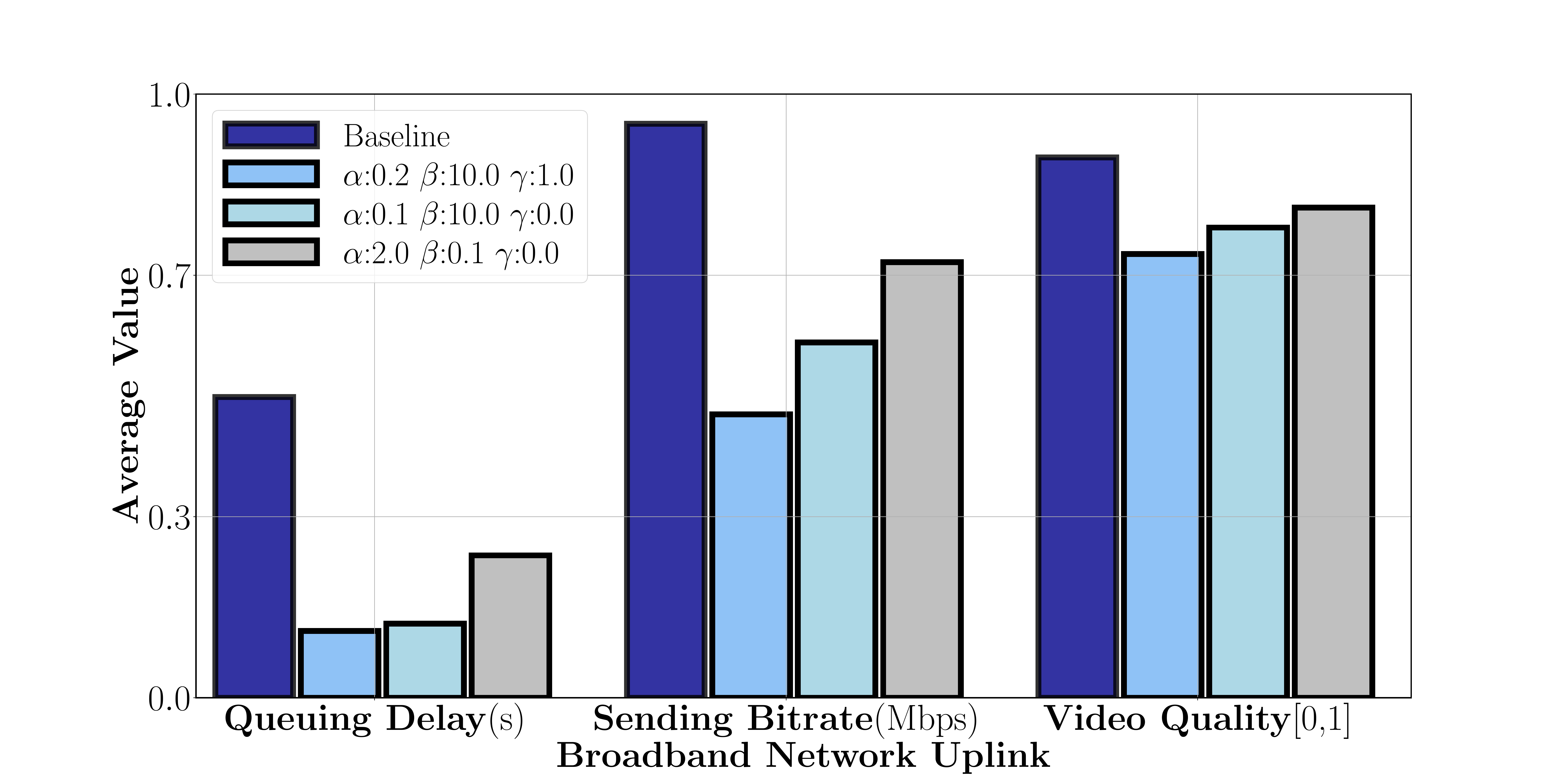}
  \end{minipage}
  \begin{minipage}{0.33\linewidth}
      \centering
      \includegraphics[width=1.0\textwidth]{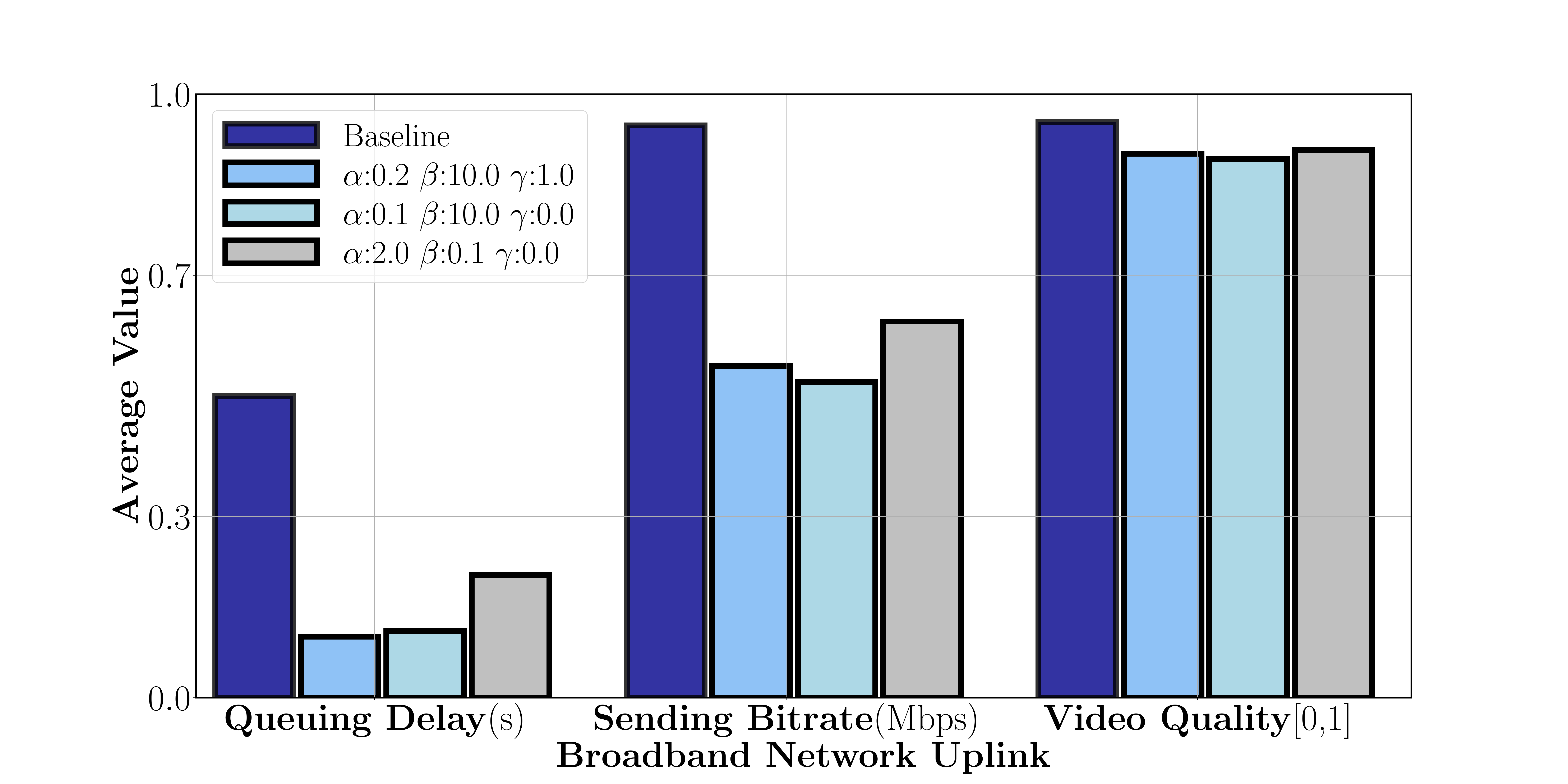}
  \end{minipage}  
  \begin{minipage}{0.33\linewidth}
      \centering
      \includegraphics[width=1.0\textwidth]{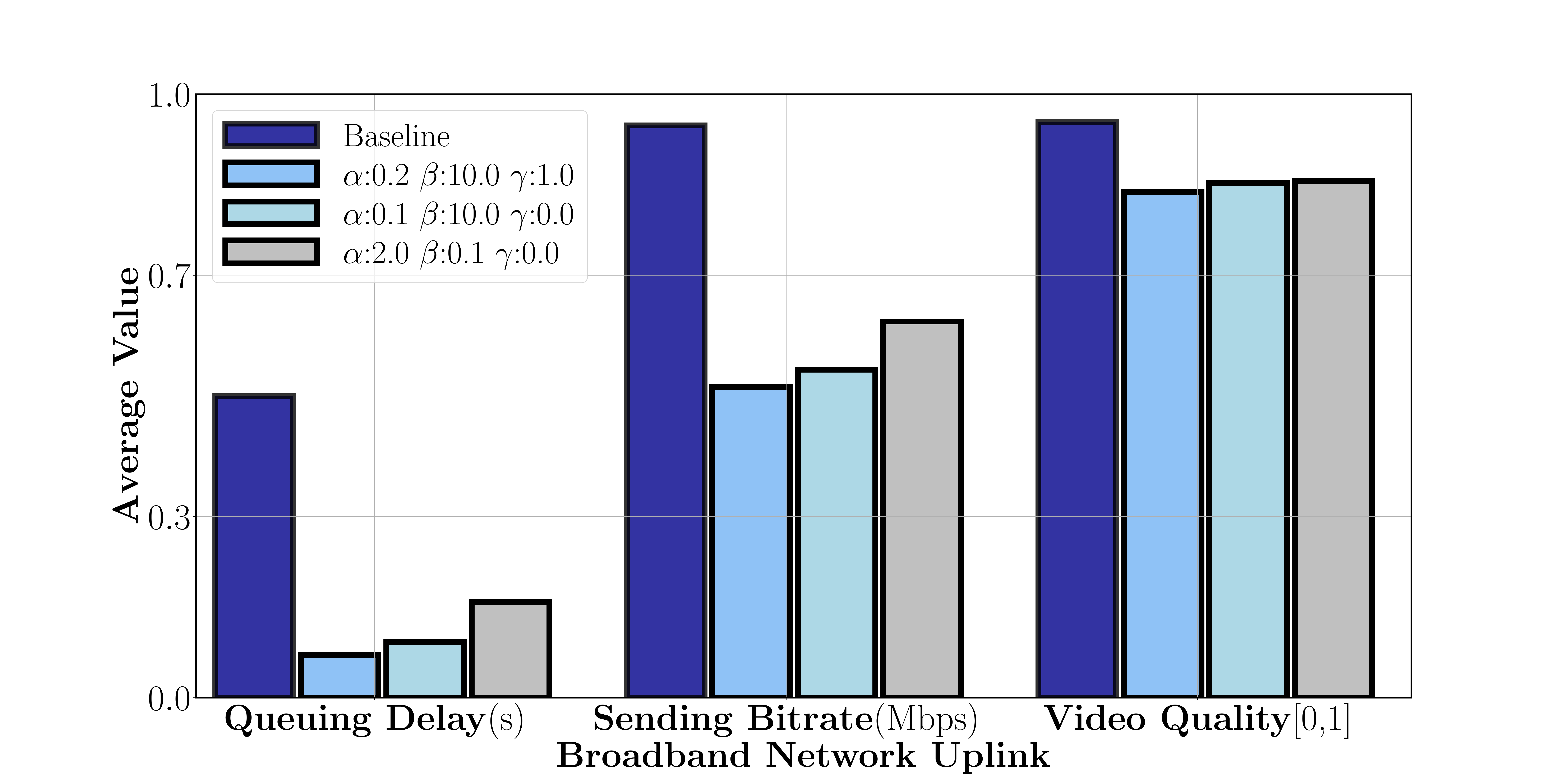}
  \end{minipage}
  \vspace{-10pt}
  \caption{Comparing QARC with different QoE and the baseline which is computed as an offline optimal value based on high video bitrate. We evaluate several QARC methods and a baseline on the \textbf{broadband network environments}. Like the process of Figure~\ref{fig:exp3}, after testing three video clips, results are shown as average queuing delay, average sending rates, and average video quality which are against the performance of the baseline value.}
  \label{fig:exp1}
  \centering
  \begin{minipage}{0.33\linewidth}
      \centering
      \includegraphics[width=1.0\textwidth]{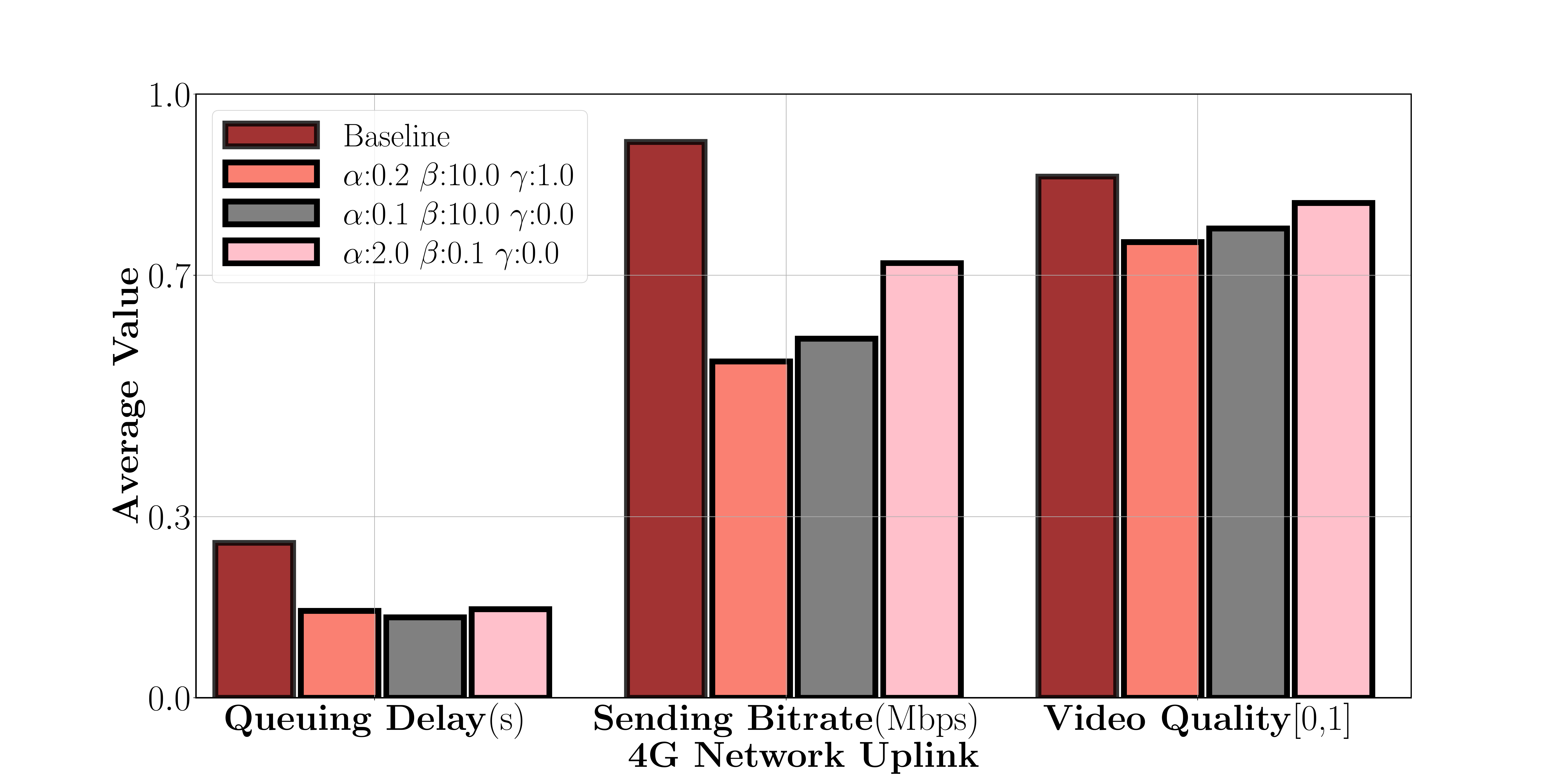}
  \end{minipage}
  \begin{minipage}{0.33\linewidth}
      \centering
      \includegraphics[width=1.0\textwidth]{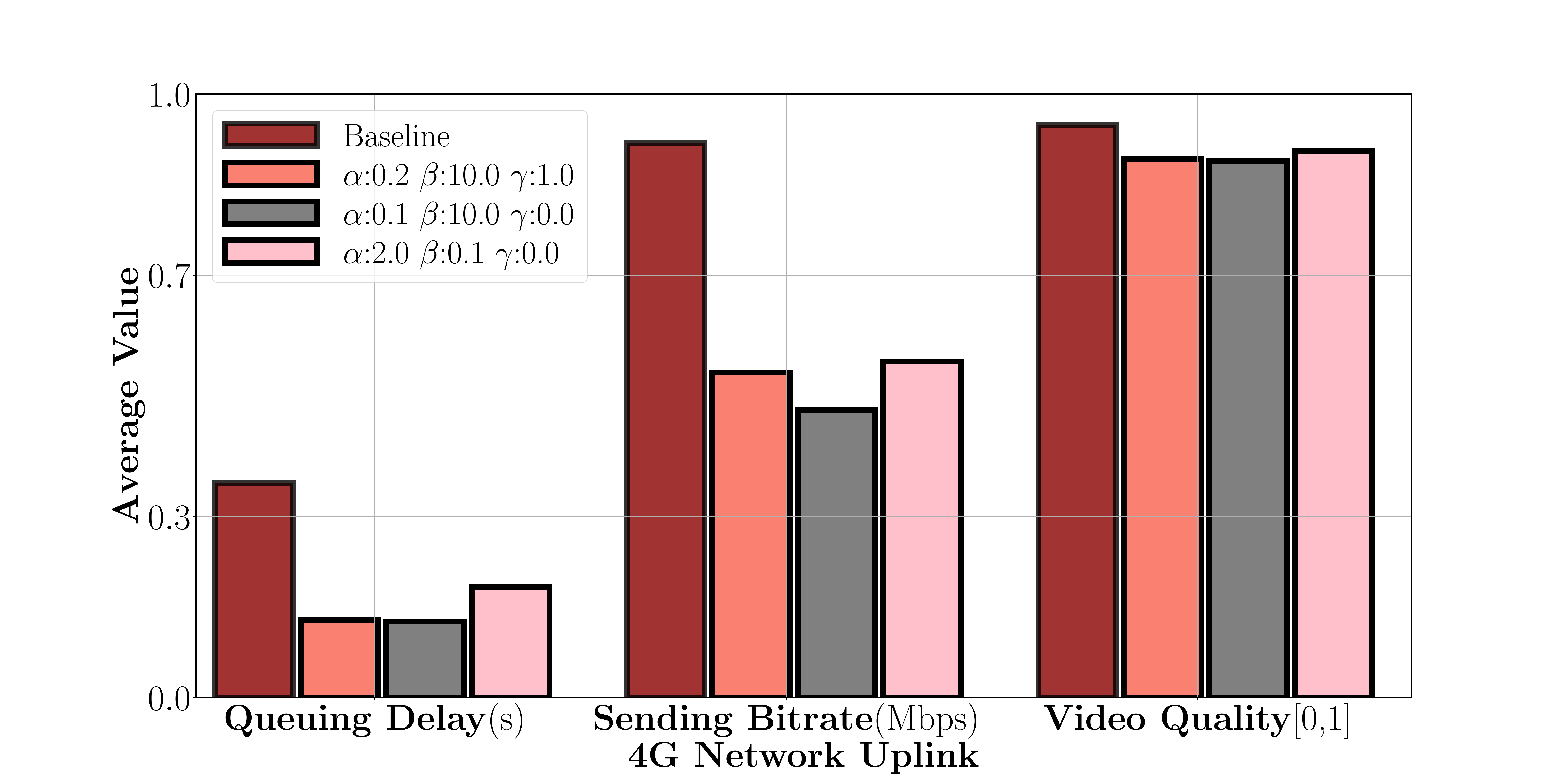}
  \end{minipage}  
  \begin{minipage}{0.33\linewidth}
      \centering
      \includegraphics[width=1.0\textwidth]{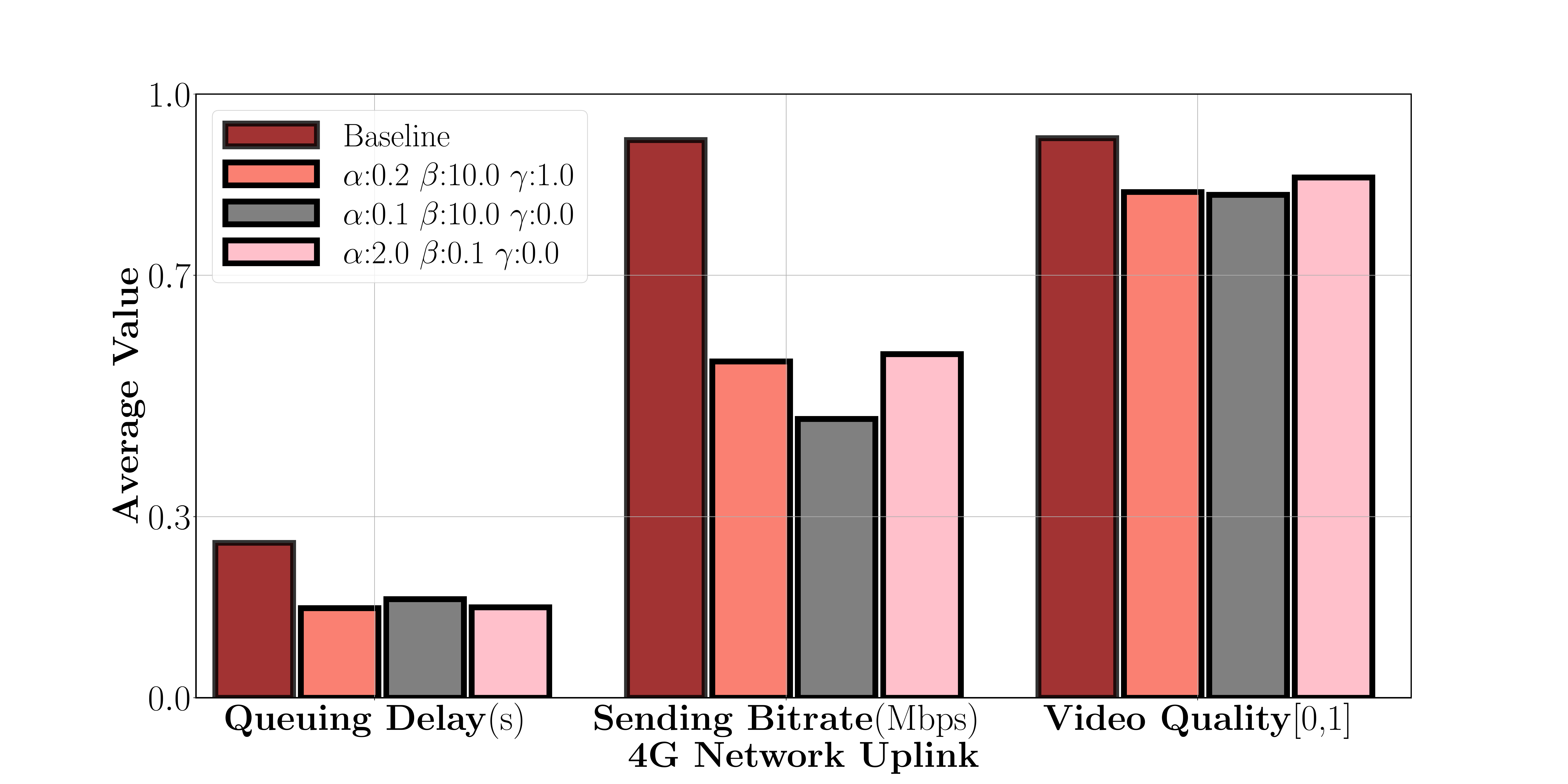}
  \end{minipage}
  \vspace{-10pt}
  \caption{Like the process of Figure~\ref{fig:exp1}, comparing QARC with different QoE and the baseline which is computed as an offline optimal value based on high video bitrate. We evaluate several QARC methods and a baseline on the \textbf{4G network environments}.}
  \label{fig:exp2}
\end{figure*}

\label{sec:VQRL_exp}
Then, we consider validating the importance of adding FFT feature into inputs. We set up two CNN models, one of them is established with FFT feature. We set sequence length $k=20$ with the same environment as the first experiment. Results are shown in Figure~\ref{fig:VQRL}(b), which implies that the CNN model with using FFT feature can provide a high reward with the improvement of about 29\% compared with the CNN model without using FFT feature.

Finally, we investigate how CNN parameters inflect output results. In our experiment, the different parameters are set as \{$k=5,c=64$\}, \{$k=10,c=64$\} and \{$k=20,c=128$\}, in which $k$ is the input sequence length and $c$ is the CNN channel size. As shown in Figure~\ref{fig:VQRL}(c), with the increase of $k$ and $c$, the performance increases. However, when we choose parameter \{ $k=20,c=128$\}, the average QoE only increases 1\% compared with parameter \{$k=10,c=128$\}, so in consideration of calculation complexity, we finally choose \{$k=10,c=64$\}. Additionally, the action space is configured as 5, which is same as the output of VQPN. During the training process, we use Adam gradient optimizer to optimize it, and the learning rate for the actor and critic is set as $10^{-4}$ and $10^{-3}$, respectively.


\textbf{Training time:} To measure the performance limitation of predicting future video quality, we profile VQPN's training process. To know when the network converges, we use early stopping method to train the neural network. Totally, training VQPN requires approximately an hour on a single GPU GTX-1080Ti.

For measuring the overhead of the neural network of VQRL, we also introduce the training process for it. To train this, we use 8 agents to update the parameters of the central agent in parallel. The neural network will converge in 22 hours, or less than 5 hours using 20 agents.\footnote{This experiment is worked on AWS with an instance in 20 CPUs and 140G RAM size.}.

\subsection{Experiments and Results}
In this section, we establish a real-time video streaming system to experimentally evaluate QARC, and use Mahimahi~\cite{netravali2015mahimahi:}, a trace-driven emulator, to simulate various network environments. Our results answer the following questions:
\begin{enumerate}
\item Comparing QARC with previously proposed approaches in different video clips, does QARC stand for the best approach? 
\item Compared with the baseline algorithm based on high video bitrate and low latency, how much improvement does QARC gain on the results?
\item How does the coefficient $\alpha$, $\beta$, and $\gamma$ affect the outcome of QARC?
\end{enumerate}

\textbf{QARC vs. Existing approaches} 
In this experiment, we evaluate QARC with existing proposed heuristic methods on several network traces which represent various network conditions by using trace-driven emulation. After running the trace for each approach, we collect the average queuing delay, average video quality and average sending rate from the receiver. We compare their performance to different video clips. In this experiment, QARC is compared with Google Hangout, a famous video conference app, Compound TCP\cite{ha2008cubic}, and Vegas~\cite{brakmo1995tcp}. As illustrated in Figure~\ref{fig:exp3}, one of the results show that QARC outperforms with existing proposed approaches, with improvements in average video quality of 18\% - 25\% and decreases in average queuing delay of 23\% - 45\%. Especially, we observe that QARC also saves the sending rate, which also performs well.  

\textbf{Video quality first vs. Bitrate first} 
In this experiment, we aim to evaluate QARC with different QoE parameters and the baseline algorithm which uses the policy based on high video bitrate.
Specifically, we compare QARC to the baseline algorithm in terms of queuing delay, the sending rate, and the video quality of the entire video session.


As shown in Figure~\ref{fig:exp1} and Figure~\ref{fig:exp2}, compared with the baseline algorithm on broadband and 4G network environments, the performance of QARC outperforms the baseline based on greedy algorithm. In the broadband network environment, despite a shrinkage in average video quality of 4\% - 9\%, QARC decreases the sending rate of 46\% to 60\% and reduces the average queuing delay \footnote{In this paper, queuing delay is regarded as self-inflicted delay, which is a lower bound on the 95\% end-to-end delay that must be experienced between a sender and receiver, given observed network behavior.~\cite{winstein2013stochastic}} from $0.5s$ to $0.04s$. It is noteworthy that if the footage of the video does not switch violently (Figure~\ref{fig:exp1}(b)), for instance, in video conference scenario, sending bitrate decreases from 51\% to 62\% while video quality reduces less than 5\%. We can also find similar results in 4G network environments. Details can be seen in Figure~\ref{fig:exp1}.

\textbf{Influence of $\alpha$,$\beta$ and $\gamma$:} Figure~\ref{fig:exp1} and Figure~\ref{fig:exp2} show the results of QARC with different initial QoE reward parameters. Unsurprisingly, initialize QoE reward with small latency coefficient $\alpha$ yield high-performance improvement over the one with a bigger $\alpha$ in wired network conditions, however, in 4G network environments, it performs a very different performance. In conclusion, there is no optimal pair can fit any network conditions. 

\section{Related Work}

\subsection{Real-time Rate Control Methods}

Traditional real-time rate control methods have been proposed and applied about two decades. These schemes are mainly classified into three types, loss-based bitrate approach, delay-based bitrate approach and model-based bitrate approach.

\textbf{Loss-based: }
Loss-based approaches such as TFRC~\cite{handley2002tcp} and rate adaptation protocol (RAP)~\cite{752152}, have been widely used in TCP congestion control, and these methods increase bitrate till packet loss occurs, which means that the actions are always late, because when packet loss occurs, latency also increases. 
Furthermore, using packet loss event as the control signal may cause its throughput to be unstable, especially in error-prone environments~\cite{geng2015delay}.

\textbf{Delay-based: }
Delay-based approaches, try to adjust sending rate to control the transmission delay, can be divided into the end-to-end delay (RTT) approaches, for example, TCP Vegas~\cite{brakmo1995tcp}; one-way delay approaches,  such as LEDBAT (Over UDP) and TCP-LP~\cite{rossi2010ledbat,kuzmanovic2006tcp}; and delay gradient approaches~\cite{carlucci2016analysis}. 


\textbf{Model-based:}
Model-based bitrate control method, such as Rebera~\cite{kurdoglu2016real}, GCC~\cite{carlucci2016analysis} and so on, they control sending bitrate based on previous network status observed including end-to-end latency, receiving rate which is measured by the receiver, and past sending bitrate, loss ratio which is measured by the sender. 

\subsection{Video Quality Metrics}
\label{sec:videoquality}
Video quality is a characteristic to measure the perceived video degradation while passing through a video transmission system. Up to now, the video quality metrics which are commonly used are shown as follows.

\textbf{PSNR:} A traditional signal quality metric~\cite{hore2010image}, which is directly derived from mean square error (MSE) or its square root (RMSE). Due to the simplicity and low complexity of its calculation, PSNR continues to be the most popular evaluation of the video quality. However, the result cannot precisely reflect the visual quality seen by human eyes.

\textbf{SSIM:} An image quality metric, submitted in 2004 by Wang et al.~\cite{1284395}. Unlike previously proposed video quality evaluation criteria, SSIM uses the structural distortion measurement instead of mean square error. Due to the consideration of the whole picture, SSMI can give a properer evaluation of the video quality experienced by users. However, SSIM is not a professional tool for video quality assessment.

\textbf{VMAF:}Video Multi-method Assessment Fusion (VMAF)~\cite{rassool2017vmaf} is an objective full-reference video quality metric which is formulated explicitly by Netflix to estimate subjective video quality based on a reference and distorted video sequence. Using machine learning techniques, VMAF provides a single output score in the range of $[0,100]$ per video frame. In general, this metric is focused on describing the quality degradation due to compression and rescaling and it is closer to users' real experience of video quality than previous schemes.

\vspace{-5pt}
\subsection{Deep Reinforcement Learning Approaches}
Deep reinforcement learning, one of the deep learning methods, aims to maximize the $reward$ of each $action$ taken by the agent in given $states$ per step. Recent years, several approaches (e.g.~
\citep{winstein2013stochastic,mao2017neural,DDASH}) have been made to optimize the network control algorithm.

\textbf{Remy:} Remy~\cite{winstein2013tcp} decides with ``a tabular method''
, and it collects experience from the network simulator with network assumptions, however, like all TCP variants, when the real network deviates from Remy`s input assumption, performance degrades. 

\textbf{Pensieve:} ~\citeauthor{mao2017neural}\cite{mao2017neural} develop a system that uses deep reinforcement learning to select bitrate for future video chunks. Unlike most of the adaptive bit rate(ABR) algorithms, Pensieve does not need any predefined rules and assumptions to make decisions, and it can automatically adjust itself to the change of network conditions. By comparing with the existing ABR algorithms, Pensieve performs very well.
\vspace{-5pt}
\section{Conclusion}
In this paper, we propose QARC, a deep-learning-based rate control algorithm in the real-time video streaming scenario. Unlike previously proposed approaches, we try to get a higher video quality with possibly lower sending rate. Due to that fixed rules cannot effectively handle the complicated scenarios caused by perplexing network conditions and various video content, we use deep reinforcement learning to select the future video bitrate, which can adjust itself automatically to the change of its inputs.
To reduce the state space of the reinforcement learning model, we derive the neural network into two parts and train them respectively. After training on a board set of network data, we explore the performance of QARC over several network conditions and QoE metrics. We find that QARC outperforms existing rate control algorithms.

\section*{Acknowledgement}
We thank the anonymous reviewers for their valuable feedback. 
The work is supported by the National Natural Science Foundation of China under Grant No. 61472204 and 61521002, Beijing Key Laboratory of Networked Multimedia No. Z161100005016051, and Key Research and Development Project under Grant No. 2018YFB1003703.


\bibliographystyle{ACM-Reference-Format}
\bibliography{sample-bibliography}

\end{document}